\documentclass[twocolumn,showpacs,amsmath,amssymb,superscriptaddress,nofootinbib]{revtex4}

\usepackage{amsfonts}
\usepackage{amsmath}
\usepackage{aas_macros}
\usepackage{graphicx}

\def\hkmsmpc{\ {\rm km\,s^{-1}\,{\it h}Mpc^{-1}}}
\def\hmpc{\ {\rm {\it h}^{-1}Mpc}}

\def\hmsun{\ {\rm M_\odot/{\it h}}}
\def\hhmpc{\ {\rm {\it h}^2Mpc^{-2}}}
\def\hhhmpc{\ {\rm {\it h}^{3}Mpc^{-3}}}
\def\hgggpc{\ {\rm {\it h}^{-3}Gpc^{3}}}

\def\hmmpc{\ {\rm {\it h}Mpc^{-1}}}

\def\la{\langle}
\def\ra{\rangle}
\def\dd{{\rm d}}
\def\ln{{\rm ln}}
\def\tr{{\rm tr}}
\def\det{{\rm det}}

\def\mathbi#1{\textbf{\em #1}}

\def\dsc{\delta_{\rm sc}}

\def\vw{{\boldsymbol\varpi}}

\def\rvh{\hat{\mathbi{r}}}
\def\zvh{\hat{\mathbi{z}}}

\def\vk{\mathbi{k}}

\def\vr{\mathbi{r}}
\def\vs{\mathbi{s}}

\def\vv{\mathbi{v}}
\def\vx{\mathbi{x}}
\def\vy{\mathbi{y}}
\def\voo{{\cal O}}
\def\vbb{{\rm B}}
\def\vcc{{\rm C}}
\def\vii{{\rm I}}
\def\vmm{{\rm M}}
\def\vpp{{\rm P}}
\def\vqq{{\rm Q}}
\def\vrr{{\rm R}}

\def\spar{s_\parallel}

\def\sper{{\bf s}_\bot}

\def\hw{\hat{W}}

\def\npk{n_{\rm pk}}
\def\bnpk{\bar{n}_{\rm pk}}

\def\vvpk{\mathbi{v}_{\rm pk}}

\def\xpk{\xi_{\rm pk}}
\def\xpd{\xi_{{\rm pk},\delta}}

\def\xlt{\xi_{\rm L}}
\def\ppk{P_{\rm pk}}
\def\ppd{P_{{\rm pk},\delta}}
\def\ppt{P_{{\rm pk},\theta_{\rm pk}}}
\def\ptt{P_{\theta_{\rm pk}}}

\def\MNRAS{{Mon.~ Not.~ R.~ Astron.~ Soc.~}}

\def\PRD{{Phys.~ Rev.~ D.~}}
\def\PRL{{Phys.~ Rev.~ Lett.~}}
\def\ApJ{{Astrophys.~ J.~}}
\def\ApJS{{Astrophys.~ J.~ Suppl.~}}
\def\ApJL{{Astrophys.~ J.~ Lett.~}}

\def\JCAP{{JCAP}}
\def\BAMS{{Bull.~Am.~Math.~Soc.}}

\def\PASJ{{Pub.~Astron.~Soc.~Jap.}}

\begin{document}

\title{Redshift space correlations and scale-dependent  stochastic
       biasing of density peaks}

\author{Vincent Desjacques} \email{dvince@physik.uzh.ch}
\affiliation{Institute for Theoretical Physics, University of Zurich, 8057 Zurich, Switzerland}
\author{Ravi K. Sheth} \email{shethrk@physics.upenn.edu}
\affiliation{Center for Particle Cosmology, University of Pennsylvania, 209 S 33rd Street, 
  Philadelphia, PA 19104, USA}


\begin{abstract}

We calculate the redshift space correlation function and the power 
spectrum of density peaks of a Gaussian random field.  Our derivation, 
which is valid on linear scales $k\lesssim 0.1\hmmpc$, is based on the 
peak biasing relation given in Desjacques [\PRD, {\bf 78}, 3503 (2008)]. 
In linear theory, the redshift space power spectrum is  
$$P^s_{\rm pk}(k,\mu) = \exp(-f^2\sigma_{\rm vel}^2\, k^2\mu^2)\,
  \left[b_{\rm pk}(k) + b_{\rm vel}(k)\,f\mu^2\right]^2\, P_\delta(k),$$
where $\mu$ is the angle with respect to the line of sight, 
$\sigma_{\rm vel}$ is the one-dimensional velocity dispersion, $f$ is
the growth rate, and $b_{\rm pk}(k)$ and $b_{\rm vel}(k)$ are 
$k$-dependent linear spatial and velocity bias factors.
For peaks, the value of $\sigma_{\rm vel}$ depends upon the functional 
form of $b_{\rm vel}$. 
When the $k$-dependence is absent from the square brackets and 
$b_{\rm vel}$ is set to unity, the resulting expression is assumed 
to describe models where the 
bias is linear and deterministic, but the velocities are unbiased.   
The peaks model is remarkable because it has unbiased velocities in 
this same sense -- peak motions are driven by dark matter flows -- but, 
in order to achieve this, $b_{\rm vel}$ is $k-$dependent.  
We speculate that this is true in general:  $k$-dependence of the 
spatial bias will lead to $k$-dependence of $b_{\rm vel}$ even if 
the biased tracers flow with the dark matter.  
Because of the $k$-dependence of the linear bias parameters, standard 
manipulations applied to the peak model will lead to $k$-dependent 
estimates of the growth factor that could erroneously be interpreted
as a signature of modified dark energy or gravity.  
We use the Fisher formalism to show that the constraint on the
growth rate $f$ is degraded by a factor of two if one
allows for a $k$-dependent velocity bias of the peak type.
Our analysis also demonstrates that the Gaussian smoothing term is part 
and parcel of linear theory. We discuss a simple estimate of nonlinear
evolution and illustrate the effect of the peak bias on the redshift
space multipoles. For $k\lesssim 0.1\hmmpc$, the peak bias is deterministic
but $k$-dependent, so the configuration space bias is stochastic and
scale dependent, both in real and redshift space.
We provide expressions for this stochasticity and its evolution.

\end{abstract}

\pacs{98.80.-k,~98.65.Dx,~95.35.+d,~98.80.Es}
\maketitle

\section{Introduction}
\label{sec:intro}

While velocities are directly measured through their Doppler
(red)shifts, accurate measurement of cosmological distances are only
available for  nearby cosmic objects, and even at these small scales
they are plagued with observational biases. Therefore, most
observational data is described in terms of redshifts,
e.g. three-dimensional (3D) galaxy surveys provide the angular
positions and redshifts of galaxies.  Redshifts differ from distances
by the peculiar velocities (deviations from pure Hubble flow) along
the line of sight.  These generate systematic differences between the
spatial distribution of data in redshift and distance (or real) space
which are commonly referred to as redshift
distortions~\cite{distortions}.  Kaiser~\cite{Kaiser1987} first
derived an expression which describes the effect of linear peculiar
motions on 3D power  spectra.   References~\cite{Fisher1995} and
\cite{Ohtaetal2004} provide  two very different derivations of this
same expression.  Whereas the  original derivation made no assumption
about the form of the density  and velocity fields, the other two
assume they are Gaussian distributed.

The Kaiser formula has been used to interpret observations of the
redshift space clustering of galaxies. The angular dependence of the
redshift distortions can be used to measure the logarithmic derivative
$f=\dd\ln D/\dd\ln a$ or growth rate \cite{Peebles1980} at multiple
redshifts and thus potentially constrain many of the dark energy or
modified gravity models (e.g. \cite{grdeviations}; for a review of
these scenarios, see \cite{DurrerMaartens2008}).  Essentially all
analyses to date assume that i) galaxies are biased tracers of the
underlying matter field, ii)  the bias is linear, local and
deterministic \cite{Kaiser1987,
Hamilton1992,Fisher1995,Coleetal1995,Scoccimarro2004}  and  iii) the
velocities of the tracers are {\em un}biased.   In fact, except on the
largest scales, the  relation between the dark matter and galaxy
fields is almost  certainly nonlinear, nonlocal, and scale dependent
\cite{Matsubara1999}.  Our main goal  in the present study is to
explore what complexities one might  expect on smaller scales where
the bias relation is more complicated,  and where the velocities may
also be biased.   We do so by investigating the impact of redshift
distortions on the  correlation function of density maxima in a
Gaussian density field.

We have chosen to study density peaks because the statistics of
Gaussian  random density~\cite{Doroshkevich1970} and velocity
fields~\cite{Gorski1988} in a cosmological context,  and of the peak
distribution in particular, has already received  considerable
attention~\cite{Kaiser1984,PeacockHeavens1985,HoffmanShaham1985,
Bardeenetal1986,Coles1989,Lumsdenetal1989,RegosSzalay1995}.  Some of
these results have been used in studies of the nonspherical  formation
of large-scale structures
~\cite{BondMyers1996,ShethMoTormen2001,Desjacques2008a,DesjacquesSmith2008}.
Others, especially from peaks theory, have been used to interpret the
abundance and clustering of rich clusters
~\cite{KaiserDavis1985,MoJingWhite1997,Cen1998,Sheth2001}. Density
peaks define a well-behaved point-process which can account for the
discrete nature of dark matter halos and galaxies.  On asymptotically
large scales, peaks are linearly biased tracers  of the dark matter
field, and this bias is scale independent
\cite{Kaiser1984,Bardeenetal1986,MoJingWhite1997}.  However, these
conclusions are based on a configuration-space  argument known as the
peak background split.  Extending the  description of peak bias to
smaller scales is more easily accomplished  by working in Fourier
space. It has been shown that peaks are linearly biased with  respect
to the mass, but this bias is $k$-dependent
~\cite{Matsubara1999,Desjacques2008b}.

The first part of this paper demonstrates that, in the large-scale
limit,  the configuration and Fourier-based approaches yield
consistent results.  This  is important, because the first (and only
other) study of the redshift  space clustering of peaks,
reference~\cite{RegosSzalay1995}, reported  that in redshift space
peaks behave very differently from the  deterministic, linear and
scale independent biased tracers investigated   in
\cite{Kaiser1987,Hamilton1992,Fisher1995,Scoccimarro2004}.   Since the
linear bias assumption that is extensively advocated to  convert large
scale redshift space measurements into information  about the
background cosmology \cite{Coleetal1995,PercivalWhite2009},  the fact
that peaks might behave very differently is potentially very worrying.
In addition, peak velocities exhibit a $k$-dependent bias even though
peaks locally flow with the dark matter  \cite{Desjacques2008b}.  This
is remarkable given that one commonly refers to such flows as  having
{\em un}biased velocities. We explain the origin of this effect and
argue that it should be a generic feature of any $k$-dependent
spatial bias model.  Again, however, peaks are remarkable because,  in
the high peak limit where their spatial bias is expected to be  linear
and scale independent, their velocity bias remains $k$-dependent.

The second part shows that, at the linear order, redshift space
distortions  for peaks can be recast in a way that retains the
simplicity of the original Kaiser formulae \cite{Kaiser1987} while
generalizing them to tracers whose linear bias is $k$-dependent.
Because the present derivation is based on a model which is supposed
to be accurate at smaller scales, we can identify an important term
which does not appear in \cite{Kaiser1987}. Furthermore, our analysis
reaches very  different conclusions from that of
~\cite{RegosSzalay1995}. Our peaks-based formula for redshift space
distortions, which  includes $k$-dependent linear bias factors for
both the density and  the velocity fields, has a rich structure.  We
hope it will serve  as a guide for what one might expect in the case
of more realistic (nonlinear, nonlocal, scale dependent) bias
prescriptions.

In the last part of this study, we use the Fisher formalism to
quantify the extent to which any $k$-dependent velocity bias of the
peak type would degrade the uncertainties on the growth rate $f$. We
also demonstrate the stochastic nature of the peak bias and discuss
its evolution with redshift. The peak biasing is interesting because,
although it is deterministic in Fourier space, it is stochastic in
real space.  A final section summarizes our findings and speculate on
some implications of the peak model.

Throughout the paper we work in the ``distant observer'' limit,  where
the line of sight is oriented along the $z$ direction.  In all
illustrative examples, we assume a flat $\Lambda$CDM cosmology with
$\Omega_{\rm m}=0.279$,  $\Omega_{\rm b}=0.0462$, $h=0.7$, $n_s=0.96$
and a present-day  normalisation $\sigma_8=0.81$
~\cite{Komatsuetal2009}.  It will also be convenient to work with
scaled velocities  $v_i\equiv \mbox{v}_i/(aHf)$, where $\mbox{v}_i$ is
the (proper) peculiar velocity,  $H\equiv d\ln a/dt$, and $f\equiv
d\ln D/d\ln a$ with $D(z)$ the linear theory growth factor.   At
$z=0.5$ this is $aHf\approx 61\hkmsmpc$.   As a result, $v_i$ has
dimensions of length.

\section{Properties of density peaks}
\label{sec:pkproperties}

We begin by reviewing some general properties of peaks in Gaussian 
random fields.  We then discuss the biasing relation which is used 
in the calculation of the redshift space correlation of density 
maxima.

\subsection{Spectral moments}

The statistical  properties of density peaks depend not only on the
underlying density field, but also on its first and second
derivatives.  We are, therefore, interested in the linear (Gaussian)
density field $\delta(\vx)$ and its first and second derivatives,
$\partial_i\delta(\vx)$ and $\partial_i\partial_j\delta(\vx)$. In this
regard, it is convenient to introduce the normalised  variables
$\nu=\delta(\vx)/\sigma_0$ and $u=-\nabla^2\delta(\vx)/\sigma_2$,
where  the $\sigma_n$ are the spectral moments of the matter power
spectrum,
\begin{equation}
\sigma_n^2 \equiv \frac{1}{2\pi^2}\int_0^\infty\!\!dk\,k^{2(n+1)}\,
  P_\delta(k,z) \hw(k,R_S)^2\;.
 \label{eq:mspec}
\end{equation} 
Here, $P_\delta(k,z)$ denotes the dimensionless power spectrum of the
linear density field at redshift $z$, and $\hw$ is a spherically
symmetric smoothing kernel of length $R_S$ (a Gaussian filter will be
adopted throughout this paper) introduced to ensure convergence of all
spectral moments.   We will use the notation $P_{\delta_S}(k,z)$ to
denote  $P_\delta(k,z)\hw(k,R_S)^2$.  The ratio $\sigma_0/\sigma_1$ is
proportional to the typical separation between zero-crossings of the
density field~\cite{Bardeenetal1986}.   For subsequent use, we also
define the spectral parameters
\begin{equation}
 \gamma_n=\frac{\sigma_n^2}{\sigma_{n-1}\sigma_{n+1}}
 \label{eq:gammas}
\end{equation}
which reflect the range over which $k^{2n+1}P_{\delta_S}(k,z)$ is
large.

We will also need the analogous quantities to $\sigma_n^2$ but for
non-zero lag:
\begin{equation}
 \xi_\ell^{(n)}(r)= \frac{1}{2\pi^2}\int_0^\infty\!\! dk\,
  k^{2(n+1)} P_{\delta_S}(k,z)\; j_\ell(kr)\;,
 \label{xielln}
\end{equation}
where $j_\ell(x)$ are spherical Bessel functions. As $\ell$ gets
larger, these harmonic transforms become increasingly sensitive to
small-scale power.

Finally, we note that the auto- and cross-correlations of the fields
$v_i$, $\delta$, $\partial_i\delta$ and $\partial_i\partial_j\delta$
can generally be decomposed  into components with definite
transformation properties under rotations.
Reference~\cite{Desjacques2008b} gives explicit expressions for the
isotropic and homogeneous linear density field.

\subsection{Smoothing scale and peak height}

The peak height $\nu$ and the filtering radius $R_S$ could in
principle be treated as two independent variables.  However, in order
to make as much connection with dark matter halos (and, to a lesser
extent, galaxies) as possible, we assume that density maxima with
height $\nu=\dsc(z)/\sigma_0(R_S)$ identified in the primeval density
field smoothed at scale $R_S$ are related to dark matter halos of mass
$M_S$ collapsing at redshift $z$,  where $\dsc(z)$ is the critical
density for collapse at $z$  in the spherical
model~\cite{GunnGott1972,PressSchechter1974}. For sake of
illustration, we will present results at $z=0.5$.  In the background
cosmology we assume, the linear critical density for (spherical)
collapse at $z=0.5$ is  $\dsc\approx 1.681$.  The  Gaussian smoothing
scale at which $\nu=1$ is $R_\star\approx 1.3\hmpc$,  so the
characteristic mass scale is  $M_\star\approx 6.5\times 10^{11}\hmsun$.

While there is a direct correspondence between  massive halos in the
evolved density field and the largest maxima  of the initial density
field, the extent to which galaxy-sized halos trace the initial
density maxima is unclear.   Therefore, we will only consider mass
scales $M_S$ significantly larger than the characteristic mass for
clustering, $M_\star$, for  which the peak model is expected to work
best.   We will present  results at redshift $z=0.5$ for two
(Gaussian)  filtering lengths, $R_S = 2.5\hmpc$ and $R_S=4\hmpc$;
these correspond  to masses $M_S = 1.9\times 10^{13}\hmsun$ and
$7.8\times 10^{13}\hmsun$, which roughly match the mean redshift and
typical mass of halos harbouring luminous red galaxies (LRGs) in the
Sloan Digital Sky Survey
(SDSS)~\cite{Mandelbaumetal2006,Kulkarnietal2007, Wake2dFSDSS2008}.
This makes  $\sigma_0/\sigma_1 \approx 3.2\hmpc$ and $4.9\hmpc$.   To
help set scales in the discussion which follows, the associated values
of $(\nu,b_\nu,b_\zeta)$ are $(2.1,1.0,16.4\hhmpc)$ and
$(2.8,2.8,43.0\hhmpc)$.    The three-dimensional velocity dispersion
of these peaks is $\sigma_{-1}^2\,(1-\gamma_0^2)$: for our two
smoothing scales, this corresponds to $(7.12\hmpc)^2$ and
$(6.66\hmpc)^2$ (recall that our velocities are in units of $aHf$, so
they have dimensions of (length)$^2$).

\subsection{Biasing}\label{sec:bias}

The large-scale asymptotics $r\to\infty$ of the two-point correlation
$\xpk(r)$ and line of sight mean streaming $[v_{12}\cdot\rvh](r)$ for
{\it discrete} local maxima of height $\nu$ can be thought of as
arising from the {\it continuous}, nonlinear bias relation
~\cite{Desjacques2008b}
\begin{align}
 \delta\npk(\vx) &= b_\nu \delta_S(\vx) - b_\zeta \nabla^2\delta_S(\vx) 
 \nonumber \\
 \vvpk(\vx) &= \vv_S(\vx)-\frac{\sigma_0^2}{\sigma_1^2}\nabla\delta_S(\vx)\;,
\label{eq:pkbiasing}
\end{align}
where $\vv_S$ is the dark matter velocity smoothed at scale $R_S$ (so
as to retain only the large-scale, coherent motion of the peak), and
the bias parameters $b_\nu$ and $b_\zeta$ are
\begin{align}
 b_\nu &=
 \frac{1}{\sigma_0}\left(\frac{\nu-\gamma_1\bar{u}}{1-\gamma_1^2}\right),
  \nonumber\\
 b_\zeta &=
 \frac{1}{\sigma_2}\left(\frac{\bar{u}-\gamma_1\nu}{1-\gamma_1^2}\right)
   = \frac{\sigma_0^2}{\sigma_1^2}\frac{\left(\nu - \sigma_0 b_\nu\right)}
   {\sigma_0}\;.
\label{eq:biases}
\end{align}
Here, $\bar u$ denotes the mean curvature of the peaks. Furthermore,
$b_\nu$ is dimensionless, whereas $b_\zeta$ has units of
(length)$^2$. Note that $b_\nu$ is precisely the amplification factor
found by~\cite{Bardeenetal1986} who neglected derivatives of the
density   correlation function (i.e. their analysis assumes
$b_\zeta\equiv 0$). We emphasize that Eq.(\ref{eq:pkbiasing}) is the
only bias relation that can account for the first order peak
correlation and mean streaming.

Strictly speaking, the bias relation (\ref{eq:pkbiasing}) is nonlocal
because of  the smoothing.  In configuration space, the peak bias
$b_{\rm pk}$ at first order could thus be defined as the convolution
\begin{equation}
  \left(b_{\rm pk}\otimes\delta\right)(\vx) \equiv
  \left(b_\nu-b_\zeta\nabla^2\right)\delta_S(\vx)\;,
  \label{eq:pkbiasr}
\end{equation}
In Fourier space, this becomes
\begin{equation}
 b_{\rm pk}(\vk) \equiv \left(b_\nu + b_\zeta k^2\right) \hw(k,R_S)
 \label{eq:pkbiask}
\end{equation}
so it has the same functional form as Eq.~(57) in  reference
\cite{Matsubara1999} who considered density extrema. Our coefficients
thus agree with those of \cite{Matsubara1999} only in the limit
$\nu\gg 1$, in which nearly all extrema are local maxima.

This bias relation is distinct from either linear~\cite{Kaiser1984} or
nonlinear~\cite{Szalay1988,FryGaztanaga1993,Coles1993} biasing
transformations of the density field for which $b_\zeta=0$.  Note in
particular that Eq.~(\ref{eq:pkbiask}) shows that {\it local} bias
schemes can generate $k$-dependent bias factors if the bias relation
involves differential operators.  Furthermore, when $\nu\gg 1$, then
$\bar u\to\gamma_1\nu$, so that  $\sigma_0 b_\nu\to \nu$ and $\sigma_2
b_\zeta\to 0$ \cite{biases}. This is clearly seen in
Fig.~\ref{fig:biases} where the biasing factors are plotted as a
function of the peak  height. Thus, the spatial bias  of the highest
peaks is expected to become scale independent, approaching the local
deterministic relation of linearly biased tracers for which there is
no $k$-dependent bias.  However, notice that the  $k$-dependence in
the velocity bias remains.  We will return to this point shortly.

\subsection{Relation to peak background split}\label{sec:pbs}

There is another route for estimating large scale bias of peaks
\cite{MoJingWhite1997} which utilizes the peak background split
argument \cite{Bardeenetal1986, ColeKaiser1989, MoWhite1996,
ShethTormen1999}.  This approach which is {\em very} different from
ours, because it is based on configuration space counts-in-cells
statistics.  In particular, it makes no mention of the bias in Fourier
space.

The large scale bias predicted by this approach is
~\cite{MoJingWhite1997}
\begin{equation}
  b_{\rm pkbs}\equiv -\frac{1}{\sigma_0\nu}
  \frac{\partial\,\ln\bigl[\bnpk(\nu)\bigr]}{\partial\,\ln\nu}
\end{equation}
where
\begin{equation}
 \bnpk(\nu)= \frac{1}{(2\pi)^2
     R_1^3}e^{-\nu^2/2}\,G_0(\gamma_1,\gamma_1\nu)
 \label{eq:npk}
\end{equation}
is the differential averaged number density of peaks in the range
$\nu$ to $\nu+d\nu$ ~\cite{Bardeenetal1986}.  Here $R_1 =
\sqrt{3}\sigma_1/\sigma_2\propto R_S$ characterises the typical radius
of density maxima, and $G_0$ is given by  setting $n=0$ in our
equation~(\ref{eq:gk}).  Therefore,
\begin{equation}
 b_{\rm pkbs} = \frac{\nu^2 + g_1}{\sigma_0\,\nu},
\end{equation}
where
\begin{equation}
 g_1 \equiv -  \frac{\partial\,\ln G_0(\gamma_1,y)}{\partial\,\ln
  y}\Biggl|_{y=\gamma_1\nu}.
\end{equation} 
Performing the derivative yields
\begin{equation}
 g_1 = -\gamma_1\nu\,
        \frac{G_1(\gamma_1,\gamma_1\nu)/G_0(\gamma_1,\gamma_1\nu)-\gamma_1\nu}
        {1-\gamma_1^2},
\end{equation} 
where $G_1$ is given by equation~(\ref{eq:gk}) with $n=1$.   However,
$G_1/G_0 \equiv \bar u$ (see the discussion immediately following
equation~\ref{eq:gk}), so
\begin{equation}
 g_1 = -\gamma_1\nu\,\left(\frac{\bar u - \gamma_1\nu}{1-\gamma_1^2}\right) 
     = -\gamma_1\nu\,\sigma_2\,b_\zeta.
\end{equation} 
The last equality follows from the definition of $b_\zeta$
(Eq.~\ref{eq:biases}). Equations~(\ref{eq:gammas})
and~(\ref{eq:biases})  eventually imply that
\begin{equation}
 -\gamma_1\nu\,\sigma_2\,b_\zeta = -\nu\,(\nu-b_\nu\sigma_0),
\end{equation} 
so 
\begin{equation}
 b_{\rm pkbs} = \frac{\nu^2 - \nu (\nu - b_\nu\sigma_0)}{\sigma_0\,\nu}
              = b_\nu.
 \label{eq:bpbs}
\end{equation}
This demonstrates that the large-scale, constant, deterministic bias
factor  returned by the peak background split approach is exactly the
same  as in our approach, when one considers scales large enough  such
that the $k$-dependence associated with the $b_\zeta$ term can be
ignored.

This is very reassuring for two reasons.  First, recall that  our
expressions for $b_\nu$ and $b_\zeta$ only agree with those given in
\cite{Matsubara1999} in the limit $\nu\gg 1$ (in which extrema are
almost certainly peaks).  The analysis above shows  that our $b_\nu$
is the appropriate generalization to lower $\nu$.   And second, the
peak background split approximation has been shown  to provide an
excellent description of large scale peak bias in simulations
\cite{MoJingWhite1997, Taruya2001}.  Since our expressions reproduce
this limit, we have confidence that our approach will provide  a good
approximation on the smaller scales where the peak-background  split
fails (i.e., where the bias $b_{\rm pk}$ becomes scale dependent).

\subsection{Power spectra and correlation functions}

Using the bias relations~(\ref{eq:pkbiasing}),  it is straightforward
to show that the real space cross- and auto-power spectrum are
\begin{align}
 \ppd(k) &= \left(b_\nu+b_\zeta\,k^2\right)\,P_\delta(k)\,\hw(k,R_S)
 \label{eq:ppd}\\
 \ppk(k) &= \left(b_\nu+b_\zeta\,k^2\right)^2\,P_{\delta_S}(k)\,.
 \label{eq:ppk}
\end{align}
We have omitted the explicit redshift and $\nu$-dependence for
brevity.  The corresponding relations for the correlation functions
are
\begin{align}
 \xpd(r) &= b_\nu\,\xi_0^{(0)}\!(r)+b_\zeta\,\xi_0^{(1)}\!(r)\,
          \label{eq:xpd}\\
 \xpk(r) &= b_\nu^2\,\xi_0^{(0)}\!(r) + 2b_\nu
          b_\zeta\,\xi_0^{(1)}\!(r) + b_\zeta^2\,\xi_0^{(2)}\!(r)\,
          \nonumber\\ &\equiv b_\xi^2(r)\,\xi_0^{(0)}(r)
          \label{eq:xpk}
\end{align} 
where the final expression defines the (scale dependent) peak bias
factor in configuration space.  As shown in ~\cite{Desjacques2008b}
(for $\xpk(r)$) and in Appendix  \ref{sec:app1} (for $\xpd(r)$), these
expressions agree with those obtained from a rather lengthy derivation
based on the peak constraint, which involves joint probability
distributions of the density field and its derivatives.  It is worth
noticing that, while expressions (\ref{eq:ppk}) and (\ref{eq:xpk}) for
the auto-power spectrum and correlation are only valid at first order
in the correlation functions $\xi_\ell^{(n)}$,  the cross-power
spectrum (\ref{eq:ppd}) and correlation (\ref{eq:xpd}) are exact to
all orders.

\subsection{Velocities}\label{sec:velb}

\begin{figure}
\resizebox{0.5\textwidth}{!}{\includegraphics{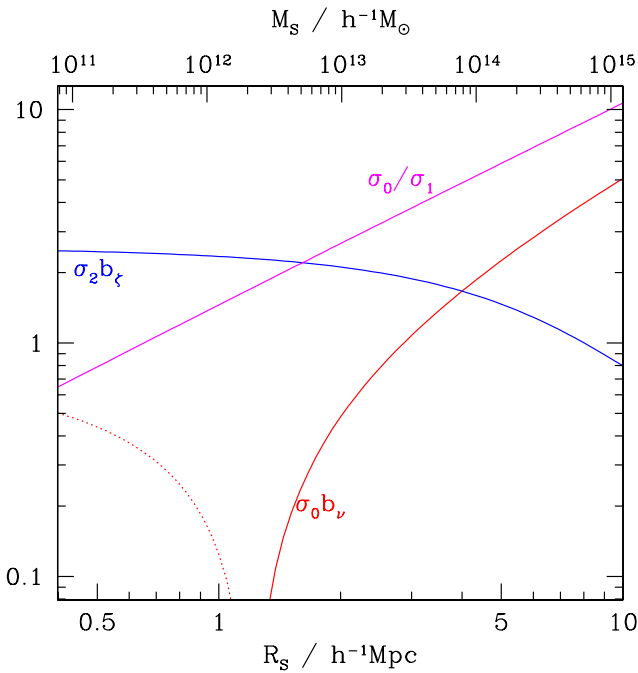}}
\caption{Bias parameters $\sigma_0 b_\nu$, $\sigma_2 b_\zeta$ and
ratio of spectral moments $\sigma_0/\sigma_1$ as a function of the
filtering scale.  Results are shown for density maxima of height
$\nu=\dsc/\sigma_0$ at redshift $z=0.5$. Dotted curves denote negative
values. The  linear spatial bias of peaks becomes scale independent in
the  limit $\nu\to\infty$. However, the $k$-dependence of the velocity
bias, which is controlled by $\sigma_0/\sigma_1$, remains and even
increases with the peak height.}
\label{fig:biases}
\end{figure}

In what follows, we will be interested in redshift space quantities,
for which the velocity field also matters.  The bias of peak
velocities  is particularly simple in Fourier space.  Taking the
divergence of Eq.~(\ref{eq:pkbiasing}), we find
\begin{equation}
  \theta_{\rm pk}(\vx) \equiv \nabla\cdot\vvpk(\vx)  =
  \nabla\cdot\vv_S(\vx)
  -\frac{\sigma_0^2}{\sigma_1^2}\nabla^2\delta_S(\vx).
\end{equation}
The linear continuity equation stipulates that  $\theta_S(\vx) \equiv
\nabla\cdot\vv_S(\vx) = -\delta_S(\vx)$,  so the result of Fourier
transforming the expression above implies that
\begin{equation}
  \theta_{\rm pk}(\vk)=\left(1 - \frac{\sigma_0^2}{\sigma_1^2}\,
  k^2\right)\hw(k,R_S)\,\theta(\vk)\equiv b_{\rm vel}(\vk)\,\theta(\vk)\;.
  \label{eq:vpkbiask}
\end{equation}
This defines the peak velocity bias factor, $b_{\rm vel}(\vk)$, which
depends on $k$ but not on $\nu$. As seen in Fig.~\ref{fig:biases}, the
ratio $\sigma_0/\sigma_1$ increases  monotonically with the filtering
scale such that, even in the limit $R_S\rightarrow\infty$
($\nu\rightarrow\infty$) where the spatial bias is linear
($b_\xi(r)\approx b_\nu$), the peak velocities remain
$k$-dependent. In general, the linear bias approximation
$\delta\npk=b_\nu\delta_S$ with unbiased velocities $\vvpk=\vv_S$ will
provide a good description of the large-scale properties of density
peaks only when $k\ll {\rm
min}[\sqrt{b_\nu/b_\zeta},\sigma_1/\sigma_0]$. For density peaks of
height $\nu\gtrsim 1$, the square root approximately is
$(\sigma_1/\sigma_0)(\nu/\sqrt{3})$. The above condition thus becomes
$k\ll\sigma_1/\sigma_0$. For the density maxima considered here,  this
implies that the $k$-independent linear bias approximation will  be
accurate for $k\ll 0.1\hmmpc$. 
 
The three-dimensional velocity dispersion of peaks is known to be
smaller than that of the dark matter
\cite{Bardeenetal1986,SzalayJensen1987,Peacocketal1987,PercivalSchafer2008}:
\begin{equation}
 \sigma^2_{\rm vpk} = \sigma_{-1}^2\, (1 - \gamma_0^2)\;.
 \label{eq:sigmavpk}
\end{equation}
Notice that Eq.~(\ref{eq:vpkbiask}) for the peak velocity bias yields
the same number,
\begin{equation}
 \sigma^2_{\rm vpk} 
  = \frac{1}{2\pi^2}\int_0^\infty\!\!dk\,P_{\delta_S}(k)\, 
    b^2_{\rm vel}(k),
 \label{eq:sigma2bsq}
\end{equation}
as it should, but that 
\begin{equation}
 \sigma^2_{\rm vpk} 
  = \frac{1}{2\pi^2}\int_0^\infty\!\!dk\,P_{\delta_S}(k)\, 
    b_{\rm vel}(k)
 \label{eq:sigma2b}
\end{equation}
also!  If we regard the integral over one power of $b_{\rm vel}$,  say
$\la b_{\rm vel}\ra$, as the peak-dark matter velocity variance  at
the same point (when smoothed on the scale of the peak), then the fact
that $\la b_{\rm vel}\ra = \la b_{\rm vel}^2\ra$ indicates that, at
the position of the peak, the velocities of the peak and the mass are
the same.  This can also be seen in the average bias relation,
Eq.~(\ref{eq:pkbiasing}): at the position of the peak the gradient of
the density vanishes  (by definition), and so $\vvpk(\vx_{\rm
pk})=\vv(\vx_{\rm pk})$.  The peak velocity dispersion is lower than
that of the mass because large scale flows are more likely to be
directed towards peaks than to be oriented randomly.  This illustrates
an important point: peaks are biased tracers which move with the dark
matter flows -- so although there is no physical bias in the
velocities, there is a  statistical bias which arises from the spatial
bias.  In the  case of peaks, the spatial bias implies that $b_{\rm
vel}$ is  $k$-dependent, and this introduces $k$-dependence into a
number  of peak-velocity statistics (we provide an explicit
calculation  of this in equation~\ref{eq:v12simple} below).    This is
almost certainly true in general:  $k$-dependence of the spatial bias
will lead to $k$-dependence  of $b_{\rm vel}$ even if the tracers flow
with the dark matter.  Note, however, that this is not a necessary
condition since, for  the highest peaks, the velocity bias remains
scale dependent even  though the spatial bias has no $k$-dependence.

The equality $\la b_{\rm vel}\ra=\la b_{\rm vel}^2\ra$ does not
uniquely constrain the velocity bias.  For instance, the choice
$b_{\rm vel}(k) = 1 - (\sigma_{-1/2}^2/\sigma_0^2)\, k$ also has  $\la
b_{\rm vel}\ra=\la b_{\rm vel}^2\ra$.  However, if we think  of the
velocity bias as a real space operator $b_{\rm vel}(\vx)$ that  maps a
vector (velocity) field onto another vector field, then for
homogeneous and isotropic random fields $b_{\rm vel}(\vx)$ {\it must}
transform as a scalar under rotations.  Hence, it must be built from
powers of the Laplacian $\nabla^2$, and this brings down a factor of
$k^2$ upon a Fourier transformation. Therefore, we generically expect
the lowest order $k$-dependence to scale as  $b_{\rm vel}(k)\equiv 1-
R_{\rm vel}^2 k^2$ (for some constant  $R_{\rm vel}$), at least for
tracers whose spatial bias relation can  be expressed as a local
mapping of the (smoothed) density and its  derivatives.

Before concluding, we emphasize that $b_{\rm pk}(k)$ and $b_{\rm
vel}(k)$ are {\it first order} bias parameters. We expect
contributions  from higher order spatial and velocity bias parameters
to become more  important as $k$  increases, but calculating them is
beyond the  scope of this paper.

\section{Redshift space clustering of density maxima}
\label{sec:zpk}

We derive three estimates of the redshift space clustering of  peaks.
The first generalizes the formulation of \cite{Kaiser1987} based on
linear theory of gravitational instability; it  furnishes a simple
estimate of the power spectrum.  The second extends the probabilistic
interpretation of \cite{Peebles1980,Fisher1995}; it  provides an
expression for the correlation function. Reference
\cite{Bharadwaj2001} has emphasized that, within the context of linear
theory, this description of the correlation function is exact whereas
that of \cite{Kaiser1987} is only approximate.   Analytic
approximations based on a probabilistic treatment lead to terms which,
upon Fourier transforming to obtain the power spectrum, are lacking
in the approach of \cite{Kaiser1987}.  This has recently been
emphasized by \cite{Scoccimarro2004}. Finally, our third estimate
shows that if one Fourier transforms at an earlier  stage in the
analysis, one obtains a slightly more intuitive expression  for the
redshift space power. We examine in detail the implications of this
approach for density peaks, despite the fact that much of this was
already done  by \cite{RegosSzalay1995}, for the reasons stated in the
Introduction.

\subsection{Simple estimate of redshift space clustering}

The redshift space coordinate (also in $\hmpc$ since velocities are in
unit of length) is given by $\vs=(\spar,\sper)$,
\begin{equation}
 \vs=\vx + f\bigl[\vv(\vx)\cdot\zvh\bigr]\zvh\;,
 \label{s:def}
\end{equation}
where $\zvh$ is the unit vector along the line of sight.    Therefore,
at the lowest order, the redshift space density contrast is related to
that  in real space by
\begin{equation}
 \delta^s(k,\mu) = \delta(k) + f \mu^2\,\theta(k)\,
 \label{eqn:pksdm}
\end{equation}
where $\mu$ is the cosine of the angle with the line of sight
\cite{Kaiser1987}.  For peaks, this becomes
\begin{align}
 \delta\npk^s(k,\mu) &= \delta\npk(\vk) + f\mu^2\,\theta_{\rm
                         pk}(\vk) \nonumber\\ &= b_{\rm
                         pk}(\vk)\,\delta(\vk) + f\mu^2\,b_{\rm
                         vel}(\vk)\,\delta(\vk) \nonumber \\ &=
                         \left[1+\frac{b_{\rm vel}(k)}{b_{\rm
                         pk}(k)}f\mu^2\right]\delta\npk(k)
 \label{eq:pksbiasing}
\end{align}
upon insertion of the peak bias relation ~(\ref{eq:pkbiasing}). Note
that $f\mu^2$ is now multiplied by a $k$-dependent factor.

Using the former relation, the calculation of the redshift space power
spectra at leading order is straightforward and yields
\begin{align}
 \ppd^{s0}(k,\mu) &= \bigl(b_{\rm pk}(k)  + \left[b_{\rm vel}(k)+
                    b_{\rm pk}(k)\right]\,f\mu^2 \nonumber\\ &\qquad
                    + \ b_{\rm vel}(k) f^2\mu^4\bigr)\,P_\delta(k)
       \label{eq:ppds}\\
 \ppk^{s0}(k,\mu) &= \left(b_\nu + b_\zeta\,k^2 + b_{\rm
   vel}(k)\,f\mu^2\right)^2\,P_\delta(k)
  \label{eq:ppks}
\end{align}
(the reason for introducing the superscript 0 will become clear
shortly). However, the corresponding expressions for the redshift
space correlations are lengthy; we provide them later in this Section.

It is conventional to write the redshift space power spectrum in
terms of the real-space one,
\begin{equation}
 \ppk^{s0}(k,\mu) =  \left[1 + \frac{b_{\rm vel}(k)}{b_{\rm
    pk}(k)}\,f\mu^2\right]^2\,\ppk(k).
 \label{Ppks0}
\end{equation}
Parameter constraints are then derived from the angular dependence
of $\ppk^s$, under the assumption of linear scale independent bias,
for which $b_\zeta=0$ and $b_{\rm vel}=1$, so the term which
multiplies  $\mu^2$ is $f/b_\nu$, and $b_\nu$ is assumed to be a
constant.   Our analysis shows that, for peaks, this prefactor is
$k$-dependent,  and it depends on peak height. The window functions
cancel out, leaving us with
\begin{align}
 \frac{b_{\rm vel}(k)}{b_{\rm pk}(k)} &\approx 
   \frac{1}{b_\nu}\left[1 - k^2 \left(\frac{\nu}{\sigma_0 b_\nu}
     \right)\frac{\sigma_0^2}{\sigma_1^2}\right] ~~ (k\ll 1) 
   \nonumber \\ 
   &\approx -\left(\frac{\nu}{\sigma_0}-b_\nu\right)^{-1} 
   \label{eq:blim}\\
   & \quad \times
   \left[1-\frac{1}{k^2}\left(\frac{\nu}{\sigma_0 b_\nu}-1\right)^{-1}
     \frac{\sigma_1^2}{\sigma_0^2}\right]~~ (k\gg 1) \nonumber \;.
\end{align}
Hence, unless care is taken, this will lead to constraints which
depend on $k$  even in the limit $\nu\gg 1$ where $b_\nu \to
\nu/\sigma_0$. 

\subsection{Probabilistic treatment}
\label{sec:prob}

In linear theory, the redshift space two-point correlation function
$\xi^s$ is related to that in real space by a convolution of the
two-point correlation function in real space, $\xi(r)$, with the
probability distribution for velocities along the line of sight
~\cite{Peebles1980,Fisher1995}:
\begin{equation}
 1+\xi^s(\spar,\sper) = \int \frac{dy\
          K(y)}{\sqrt{2\pi}f\sigma_{12}(r)}
          \exp\left[-\frac{(\spar-y)^2}{f^2\sigma_{12}^2(r)}\right]\;,
 \label{eq:kfisher}
\end{equation}
where
\begin{align}
 K(y) &= 1 + \xi(r) + \left(\frac{y}{r}\right)
          \frac{v_{12}(r)}{\sigma_{12}(r)}\left(\frac{\spar-y}{f\sigma_{12}(r)}
          \right) \nonumber\\ & \quad - \frac{1}{4}
          \left(\frac{y}{r}\right)^2
          \frac{v^2_{12}(r)}{\sigma^2_{12}(r)} \left[1 -
          \left(\frac{\spar-y}{f\sigma_{12}(r)}\right)^2\right].
\end{align}
Here, $v_{12}(r)$ and $\sigma_{12}(r)$ are the mean and dispersion of
the pairwise velocity distribution of pairs separated by $r$ in
real-space (note that $r^2 = y^2 + \sper^2$).   As emphasized
by~\cite{Bharadwaj2001}, within the context of linear  theory and the
plane-parallel approximation, this expression is exact. This
formulation is usually referred to as the ``streaming'' model. It
should be noted that random pairs in real space are mapped to real
space differently at different separation $r$ because the pairwise
velocity distribution depends on scale \cite{Scoccimarro2004}.

Equation~(\ref{eq:kfisher}) can be generalized to give $1 + \xpk^s$,
the redshift space correlation function of peaks, simply by replacing
$v_{12}$ and $\sigma_{12}$ with the expressions appropriate for peaks
\cite{RegosSzalay1995}. At first order, these are
\begin{align}
v_{12}(r,\mu) &= \left[1+\xpk\right]^{-1}  \times \Biggl[2
 b_\nu\left(\frac{\sigma_0^2}{\sigma_1^2}\xi_1^{(1/2)}
 -\xi_1^{(-1/2)}\right)\Biggr. \nonumber \\ &\quad\Biggl. +  2 b_\zeta
 \left(\frac{\sigma_0^2}{\sigma_1^2} \xi_1^{(3/2)}
 -\xi_1^{(1/2)}\right)\Biggr]\,L_1(\mu)\;,
 \label{eq:v12pk}
\end{align}
\begin{align}
\sigma_{12}^2(r,\mu) &=
 \Biggl[\frac{2}{3}\left(1-\gamma_0^2\right)\sigma_{-1}^2
 +\frac{2}{3}\frac{\sigma_0^2}{\sigma_1^2}\left(2\xi_0^{(0)}
 -\frac{\sigma_0^2}{\sigma_1^2}\xi_0^{(1)}\right)\Biggr. \nonumber \\
 & \quad \Biggl. -\frac{2}{3}\xi_0^{(-1)}\Biggr]
 -\frac{4}{3}\Biggl[\frac{\sigma_0^2}{\sigma_1^2}\left(2\xi_2^{(0)}
 -\frac{\sigma_0^2}{\sigma_1^2}\xi_2^{(1)}\right)-\xi_2^{(-1)}\Biggr]
 \nonumber \\
 & \quad\times L_2(\mu)\;,
 \label{eq:sig12pk}
\end{align} 
where $\mu=\rvh\cdot\zvh$ is the cosine of the angle between the line
of separation and the line of sight, and the $L_\ell(\mu)$ are
Legendre Polynomials \cite{v12error}.  

Appendix  \ref{sec:app1} demonstrates that Eq.~(\ref{eq:v12pk}) 
exactly reproduces the result of a lengthy derivation based on the 
peak constraint.  Note however, that it can be derived simply 
from setting 
\begin{equation}
 v_{12}(r,\mu) \equiv
 \frac{\langle (1 + \delta n_{\rm pk,1})(1 + \delta n_{\rm pk,2})
         ( \mathbi{v}_{\rm pk,1} - \mathbi{v}_{\rm pk,2})\cdot \zvh\rangle}
      {\langle (1 + \delta n_{\rm pk,1})(1 + \delta n_{\rm pk,2})\rangle},
 \label{eq:v12simple}
\end{equation}
where the subscripts 1 and 2 indicate positions separated by
$\vr$, and the average is over all peak pairs with separation $r$.  
The correspondence with equation~(\ref{eq:v12pk}) can be seen by 
noting that in $k$-space, the spatial bias from $\delta\npk$ 
is the sum of two terms, one of which is proportional to $k^2$ 
(equation~\ref{eq:pkbiask}) and the velocity bias 
(equation~\ref{eq:vpkbiask}) introduces additional $k^2$ 
terms which come with factors of $(\sigma_0/\sigma_1)^2$.  
Each additional factor of $k^2$ changes $\xi_1^{(n)}$ to $\xi_1^{(n+1)}$.

The first term on the right-hand side of Eq.~(\ref{eq:sig12pk}) is 
twice the (one-dimensional) velocity dispersion of peaks; recall that 
it is reduced by a factor of $1-\gamma_0^2$ relative to that of the 
dark matter (see Eq.~\ref{eq:sigmavpk}).  

\subsubsection{Approximating the integral}

When $\sigma_{12}\ll \spar$, then the Gaussian term in the expression
above will be sharply peaked around $y=\spar$.   Expanding $\xi$,
$v_{12}$ and $\sigma_{12}$ about their redshift  space values yields
\begin{equation}
 \xi^s\approx \xi - fv_{12}' + \frac{1}{2}f^2\sigma_{12}^{2''} +
                  \frac{1}{2}f^2\xi''\sigma_{12}^2\mid_\infty\;,
\label{eq:xis}
\end{equation}
where all quantities in the right hand side are evaluated at $\vs$ and
primes denote derivatives with respect to $\spar$ (recall that  $s^2 =
\spar^2 + \sper^2$). Eq.~(\ref{eq:xis}) describes the large-scale
limit of the  redshift space correlation function, in which
derivatives of the real space correlation and pairwise moments
(i.e. the  distortions) are small \cite{Scoccimarro2004}.  When
applied to dark matter  rather than density peaks, the Fourier
transform of the first three terms on  the right-hand side yields
Kaiser's formula~\cite{Fisher1995,Bharadwaj2001}.  The fourth term
arises because the pairwise velocity dispersion does  not vanish even
in the large scale limit~\cite{Scoccimarro2004}.  We show below that
Fourier transforming the analogous terms for peaks gives
Eq.~(\ref{eq:ppks}).

The derivatives of $\xpk(\vs)$, $v_{12}(\vs)$ and
$\sigma_{12}^2(\vs)$ with respect to the line of sight distance
$\spar$  can be evaluated using $ds/d\spar=\mu$ and
$d\mu/d\spar=(1-\mu^2)s^{-1}$,  which follow from the fact that $s^2 =
\spar^2 + s_\perp^2$.  The following relations are useful:
\begin{align}
  \frac{d^2\xi_0^{(n)}}{d\spar^2} &=
  \frac{2}{3}\xi_2^{(n+1)}L_2(\mu)-\frac{1}{3}\xi_0^{(n+1)},\nonumber \\
  \frac{d}{d\spar}\left[\xi_1^{(n)}L_1(\mu)\right] &=
  -\frac{2}{3}\xi_2^{(n+1/2)} L_2(\mu)+\frac{1}{3}\xi_0^{(n+1/2)}
  \nonumber \\ 
  &\quad\times\frac{d^2}{d\spar^2}\left[\xi_2^{(n)}L_2(\mu)\right]
  \nonumber \\
  &=\frac{12}{35}\xi_4^{(n+1)}L_4(\mu)-\frac{11}{21}\xi_2^{(n+1)}L_2(\mu)
  \nonumber \\
  &\quad +\frac{2}{15}\xi_0^{(n+1)}\;.
\end{align}
As a rule, terms in $\xi_\ell^{(n)}$ appear always multiplied by the
Legendre polynomial of order $\ell$. The lowest even polynomials are
$L_0(\mu)=1$, $L_2(\mu)=(3\mu^2-1)/2$ and
$L_4(\mu)=(35\mu^4-30\mu^2+3)/8$.  The calculation of the redshift
space correlation of peaks $\xpk^s(s,\mu)$ is now straightforward.
Adding all terms together, we find
\begin{widetext}
\begin{align}
\label{eq:xispk}
 \xpk^s(s,\mu) &= \frac{8}{35}f^2\left(\xi_4^{(0)}
 -2\frac{\sigma_0^2}{\sigma_1^2}\xi_4^{(1)}+\frac{\sigma_0^4}{\sigma_1^4}
 \xi_4^{(2)}\right) L_4(\mu)
 +\left\{\frac{4}{3}f\left[b_\nu\left(\frac{\sigma_0^2}{\sigma_1^2}
 \xi_2^{(1)}-\xi_2^{(0)}\right)+b_\zeta\left(\frac{\sigma_0^2}{\sigma_1^2}
 \xi_2^{(2)}-\xi_2^{(1)}\right)\right]\right. \nonumber \\ 
 & \quad\left. -\frac{4}{7}f^2\left(\xi_2^{(0)}
 -2\frac{\sigma_0^2}{\sigma_1^2}\xi_2^{(1)}+\frac{\sigma_0^4}{\sigma_1^4}
 \xi_2^{(2)}\right)+\frac{2}{9}f^2\left(1-\gamma_0^2\right)\sigma_{-1}^2
 \left(b_\nu^2\xi_2^{(1)}+2 b_\nu b_\zeta\xi_2^{(2)}+b_\zeta^2\xi_2^{(3)} 
 \right)\right\}L_2(\mu)+ b_\nu^2\xi_0^{(0)} \nonumber \\
 &\quad +2 b_\nu b_\zeta\xi_0^{(1)}+b_\zeta^2\xi_0^{(2)}
 -\frac{2}{3}f\left[b_\nu\left(\frac{\sigma_0^2}{\sigma_1^2}\xi_0^{(1)}
 -\xi_0^{(0)}\right)+b_\zeta\left(\frac{\sigma_0^2}{\sigma_1^2}\xi_0^{(2)}
 -\xi_0^{(1)}\right)\right]
 +\frac{1}{5}f^2\left(\xi_0^{(0)}-2\frac{\sigma_0^2}{\sigma_1^2}
 \xi_0^{(1)}+\frac{\sigma_0^4}{\sigma_1^4}\xi_0^{(2)}\right)\nonumber \\
 &\quad -\frac{1}{9}f^2\left(1-\gamma_0^2\right)\sigma_{-1}^2
 \left(b_\nu^2\xi_0^{(1)}+2 b_\nu b_\zeta\xi_0^{(2)}+b_\zeta^2\xi_0^{(3)}
 \right)\;.
\end{align}
As can be seen, there are harmonics up to $\xi_\ell^{(3)}(s)$ which
arise from the second derivative $\xi''(\vs)$ in Eq.~(\ref{eq:xis}).
These terms  are significant only at distances less than a  few
smoothing radii and across the baryon acoustic feature where the
density correlation $\xi_0^{(0)}$ changes rapidly
~\cite{Desjacques2008b}.  Furthermore, terms linear in $f$ arise only
from the derivative of  the pairwise velocity, $-f v_{12}'(\vs)$.

The redshift space power spectrum $\ppk^s(k,\mu)$ in this
approximation  is obtained simply by Fourier transforming
Eq.~(\ref{eq:xispk}).   For the sake of completeness,
\begin{align}
 \ppk^s(k,\mu) &=  \frac{8}{35} {\cal B}^2(k) L_4(\mu)\, P_{\rm
   pk}(k) + \left[\frac{4}{3}\,{\cal B}(k)  + \frac{4}{7}{\cal B}^2(k)
   - \frac{2}{9}f^2 k^2\left(1-\gamma_0^2\right)\sigma_{-1}^2  \right]
   L_2(\mu)\,  P_{\rm pk}(k) \nonumber \\ & \quad + \left[1 +
   \frac{2}{3}{\cal B}(k) +\frac{1}{5} {\cal B}^2(k) -\frac{1}{9}f^2
   k^2\left(1-\gamma_0^2\right)\sigma_{-1}^2  \right]\, P_{\rm pk}(k)
   \nonumber \\ &= \Bigl[1 + {\cal B}(k)\,\mu^2\Bigr]^2 \, P_{\rm
   pk}(k)  - \frac{k^2\mu^2}{3}\,f^2\,
   \left(1-\gamma_0^2\right)\sigma_{-1}^2 \, P_{\rm pk}(k)\;,
\label{eq:pkspk}
\end{align}
\end{widetext}
where the linear redshift distortion parameter
\begin{equation}
 {\cal B}(k)\equiv f \frac{b_{\rm vel}(k)}{b_{\rm pk}(k)}
\end{equation}
is scale dependent.  Recall that $b_{\rm pk}(k)$ and $b_{\rm vel}(k)$
were defined in equations~(\ref{eq:pkbiask}) and~(\ref{eq:vpkbiask}).
Thus, except for the second term in the last equality, the above
result  exactly matches our simple estimate, Eq.~(\ref{eq:ppks}).

Notice especially that, for linearly biased tracers, redshift space
distortions are used to estimate $\beta = f/b$.  The analogous
quantity for peaks, ${\cal B}(k)$, is $k$-dependent.  We will consider
the implications of this in the next section.

\subsubsection{A different approximation}

A more intuitive approximation to the exact result that is reached
upon performing the integral in Eq.~(\ref{eq:kfisher}) can be
obtained by Fourier transforming it in the first place. We write
\begin{align}
 \exp\left(-{\rm i}\vk\cdot\vs\right) &= 
                   \exp\left(-{\rm i}\vk_\perp\cdot\sper\right)\,
                   \exp\left(-{\rm i} k_\parallel y\right)\nonumber\\ 
		   & \quad\times\ 
		   \exp\bigl[-{\rm i} k_\parallel\left(\spar-y\right)\bigr]\\ 
		   &=
                   \exp\left(-{\rm i}\vk\cdot\vr\right) \,\exp\bigl[-{\rm i}
                   k_\parallel\left(\spar-y\right)\bigr], \nonumber
\end{align}
and then rearrange the order of the integrals so that the integration
over $\spar-y$ is done first. Next, we use the fact that
\begin{align}
 \int\! dt\,{\rm e}^{-t^2/2}\,{\rm e}^{-{\rm i}kt} &= {\rm
   e}^{-k^2/2} \nonumber\\ \int\! dt\,t\,{\rm e}^{-t^2/2}\,{\rm
   e}^{-{\rm i}kt} &=  -{\rm i}k\,{\rm e}^{-k^2/2} \\ \int\!
   dt\,t^2\,{\rm e}^{-t^2/2}\,{\rm e}^{-{\rm i}kt} &=  (1 -
   k^2)\,{\rm e}^{-k^2/2}, \nonumber
\end{align} 
to express the result of the integral over $\spar-y$ as $\exp[-f^2
k_\parallel^2\sigma_{12}^2(r)/2]$ times other factors.  Finally, we
recast this term as  $\exp[-f^2 k_\parallel^2\sigma_{12}^2(\infty)/2]$
times $\exp[-f^2
k_\parallel^2(\sigma_{12}^2(r)-\sigma_{12}^2(\infty))/2]$ which for
small $k_\parallel$ is approximately $\exp[-f^2
k_\parallel^2\sigma_{12}^2(\infty)/2]\times [1 - f^2
k_\parallel^2(\sigma_{12}^2(r)-\sigma_{12}^2(\infty))/2]$.   Thus, we
generically expect the redshift space power spectrum to take the form
$\exp[-f^2 k_\parallel^2\sigma_{12}^2(\infty)/2]$ times other factors.
A little algebra shows that, to leading order, these factors are
precisely those given by equation~(\ref{eq:ppks}),  giving
\begin{equation}
 \ppk^s(k,\mu) = \exp\Bigl[-f^2 k^2\sigma_{\rm
 vel}^2\mu^2\Bigr]\,\ppk^{s0}(k,\mu).
 \label{eq:rs95correct}
\end{equation}  
Here, $\ppk^{s0}$ is given by Eq.~(\ref{Ppks0}). We have also used the
fact that, except in pathological cases,  the pairwise dispersion at
very large separation is simply twice the one-dimensional velocity
dispersion of single particles,  $\sigma_{\rm vel}$ ($=\sigma_{\rm
vpk}/3$  for peaks), in units of $aHf$. Our notation is purposely kept
general  to emphasize that these results apply to any tracers of the
linear  density field.

Our equation~(\ref{eq:rs95correct}) corrects a number of important
errors in previous analyses \cite{RegosSzalay1995, mistake}.  In
addition, expanding the Gaussian smoothing term shows the origin  of
the extra terms identified in the previous subsection (those
highlighted by \cite{Scoccimarro2004}).   Finally, the form of our
expression reflects the fact that the associated  correlation function
can be written as a convolution of the  original expression
$\xpk^{s0}$ (which is the Fourier transform of Eq.~\ref{eq:ppks}) with
a Gaussian in the line of  sight direction:
\begin{equation}
 \xpk^s(\sper,\spar) = \int_{-\infty}^{+\infty}\!\!d\spar'\,
      G\left[\spar',\sigma_{12}(\infty)\right]\,
      \xpk^{s0}(\sper,\spar + \spar') \;.
\end{equation}
Now the meaning of our notation should be clear:  the superscript $0$
refers to the limit in which the dispersion of the Gaussian  smoothing
term is vanishingly small.   We note that this form for $\xi^s$ was
shown to be appropriate for  the dark matter, without using any
Fourier-space  analysis~\cite{Bharadwaj2001}; our analysis
demonstrates  that it carries  through for peaks as well.   The
interesting subtlety brought by density peaks is that the amplitude
of the damping term $\sigma_{\rm vel}$ is related to the form of
$b_{\rm vel}$ (Eq.~\ref{eq:sigma2bsq}).

The functional form of our equation~(\ref{eq:rs95correct}) has been
studied previously in the context of modelling nonlinear corrections
to the redshift space power of linearly biased tracers
\cite{Coleetal1995}, although there the assumption was  that $b_{\rm
pk}$ is constant and $b_{\rm vel}$ is unity.  We will  discuss the
effects of nonlinearities shortly.  For completeness  here, we simply
borrow all that previous analysis to show the  effect that the
Gaussian smoothing has on the Fourier space multipoles.  These can
be written as
\begin{align}
 \frac{{\cal P}_0^s(k)}{\ppk(k)} &= A_0(\kappa) +
    \frac{2}{3}A_1(\kappa){\cal B}(k)  + \frac{1}{5}A_2(\kappa) {\cal
    B}^2(k)\label{eq:p0}\;, \\ \frac{{\cal P}_2^s(k)}{\ppk(k)} &=
    \frac{5}{2}[A_1(\kappa) -
    A_0(\kappa)]+\left[3A_2(\kappa)-\frac{5}{3}A_1(\kappa)
    \right]{\cal B}(k)\nonumber\\ & \quad
    +\left[\frac{15}{14}A_3(\kappa)-\frac{1}{2}A_2(\kappa)\right]{\cal
    B}^2(k)\label{eq:p2}\;,
\end{align}
and
\begin{align}
  \frac{{\cal P}_4^s(k)}{\ppk(k)} &=
    \frac{63}{8}A_2(\kappa)-\frac{45}{4}A_1(\kappa)+\frac{27}{8}A_0(\kappa)
    \label{eq:p4} \\ & \quad +
    \left[\frac{45}{4}A_3(\kappa)-\frac{27}{2}A_2(\kappa)
      +\frac{9}{4}A_1(\kappa)
    \right]{\cal B}(k) \nonumber \\ & \quad +
    \left[\frac{35}{8}A_4(\kappa)
      -\frac{135}{28}A_3(\kappa)+\frac{27}{40}A_2(\kappa)
    \right]{\cal B}^2(k) \nonumber \;,
\end{align}
where $\kappa \equiv f k\sigma_{\rm vel}$ and the coefficients
$A_\ell(\kappa)$ are recursively defined as
\begin{align}
\label{eq:acoeff}
  A_0(\kappa) &= \frac{\sqrt{\pi}}{2} \frac{{\rm
                  erf}(\kappa)}{\kappa} \approx 1 - \frac{\kappa^2}{3}
                  \\ A_\ell(\kappa) &=
                  \frac{(2\ell+1)}{2\kappa^2}\left(A_{\ell-1}(\kappa)
                  -{\rm e}^{-\kappa^2}\right) \approx 1 -
                  \frac{(2\ell+1)}{(2\ell+3)}\kappa^2 \nonumber
\end{align}
The final approximations assume $\kappa\ll 1$.  Thus, the lowest
order corrections to $P^{s0}_{0,{\rm pk}}/\ppk$ and  $P^{s0}_{2,{\rm
pk}}/\ppk$ are proportional to $-k^2$.

\subsection{Nonlinear evolution}
\label{sub:nonlinear}

There are four reasons why nonlinear evolution will act to change the
expressions above \cite{Shethetal2001,Scoccimarro2004}: one is related
to the change in the bias parameters, and the three others have to  do
with the effect of peculiar velocities. Gravitational motions are
expected to relate a scale independent, deterministic linear bias
parameter in the initial (Lagrangian) field to the evolved (Eulerian)
bias according to $b_\nu^{\rm Eul}=1+b_\nu$ \cite{MoWhite1996}.
Ignoring the fact that $b_\zeta$ might also evolve, we follow common
practice and assume $b_{\rm pk}^{\rm Eul}\equiv 1 + b_\nu + b_\zeta
k^2$, even though we suspect that $b_{\rm pk}^{\rm Eul}\equiv b_{\rm
vel}  + b_\nu + b_\zeta k^2$ (This issue will be thoroughly explored
in a  forthcoming paper).  Note that we have omitted the smoothing
window for brevity, but it effectively makes little difference at
scales $k^{-1}\gg R_S$.  Regarding the peak motions, we first assume
that virial velocities within peaks or  halos will increase
$\sigma_{12}(\infty)$; these are responsible for the
fingers-of-god~\cite{Jackson1972,Hamilton1998} seen in galaxy surveys.
Secondly, halo/peak motions may not closely follow linear theory, but
this effect is expected to be less dramatic \cite{ShethDiaferio2001}.
Finally, the real space power spectrum will also be modified as a
result of the linear theory motions
\cite{Bharadwaj1996,CrocceScoccimarro08,Matsubara2008a,Matsubara2008b}.

For reasons we describe below, nonlinear effects may be approximated
by setting
\begin{equation}
 \ppk^s(k,\mu) = \ppk^{s0}(k,\mu)\,V_{\rm ql}(k,\mu^2)\,V_{\rm
 vir}(k,\mu^2)
 \label{eq:nonlin}
\end{equation}  
where $\ppk^{s0}$ is given by Eq.~(\ref{Ppks0}) with $b_{\rm pk}$
replaced by its nonlinear version. The filters $V_{\rm ql}$ and
$V_{\rm vir}$  are supposed to reflect the quasi-linear and virial
corrections to the {\it non-damped}  linear theory expression,
respectively.  The exact functional form of $V_{\rm ql}(k,\mu^2)$
depends upon the distribution of pairwise velocities.  However,
motivated  by results from perturbation theory
\cite{EisensteinSeoWhite07,Matsubara2008a,Matsubara2008b}, we set
\begin{equation}
 V_{\rm ql}(k,\mu^2) = \exp\Bigl[-k^2\sigma_{\rm vel}^2(1-\mu^2)
 -k^2\sigma_{\rm vel}^2(1+f)^2\mu^2\Bigr]\;.
\label{eq:damping}
\end{equation}
This takes into account both the smearing of linear power caused by
linear theory displacements and the damping due to the linear pairwise
velocity dispersion. In principle, the reduction in linear power
should be somewhat mitigated  by the addition of nonlinear
mode-coupling terms of the  sort discussed by
\cite{CrocceScoccimarro08}. However, we will ignore  these terms in
what follows.  The last multiplicative factor  $V_{\rm vir}$ accounts
for  the damping of redshift  space power due to nonlinear virial
motions within halos (assumed uncorrelated with the large-scale
flows).  If the mass range is small (i.e. if the peaks cover a  small
range in $\nu$), then $V_{\rm vir} = \exp(-k^2\mu^2\sigma_{\rm
vir}^2)$, where  $\sigma_{\rm vir}$ depends on the halo or peak mass,
should be a good approximation. If the mass range is broad, then an
exponential distribution may be more appropriate \cite{Sheth1996}, 
leading to $V_{\rm nl} = [1 + k^2\sigma_{\rm vir}^2\,\mu^2]^{-2}$.  
Removing fingers-of-god from a survey \cite[e.g.][]{TegmarkSDSS2006} 
is equivalent to setting $\sigma_{\rm vir}\to 0$ or $V_{\rm vir}\to 1$.

\begin{figure*}
\resizebox{0.48\textwidth}{!}{\includegraphics{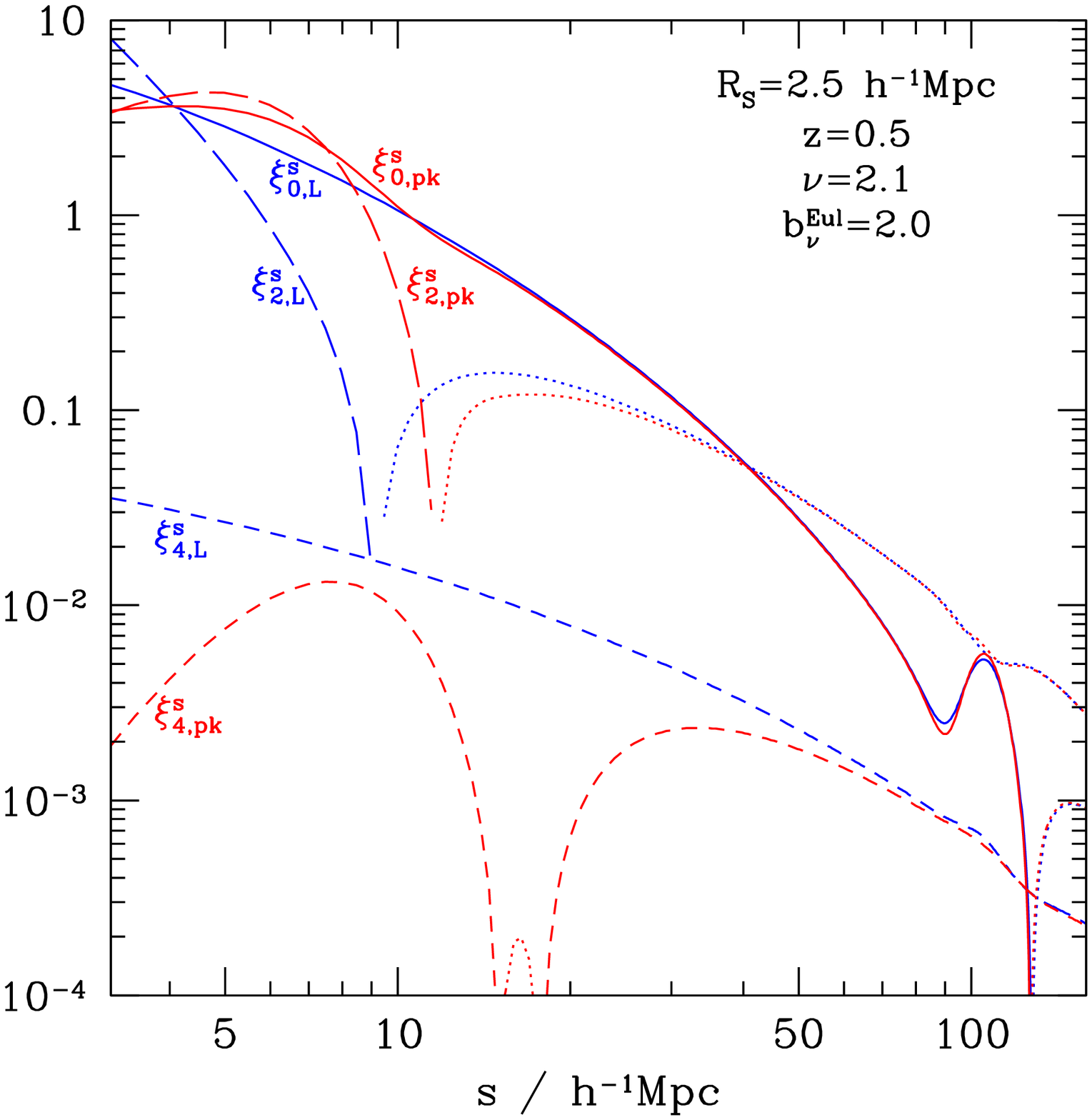}}
\resizebox{0.48\textwidth}{!}{\includegraphics{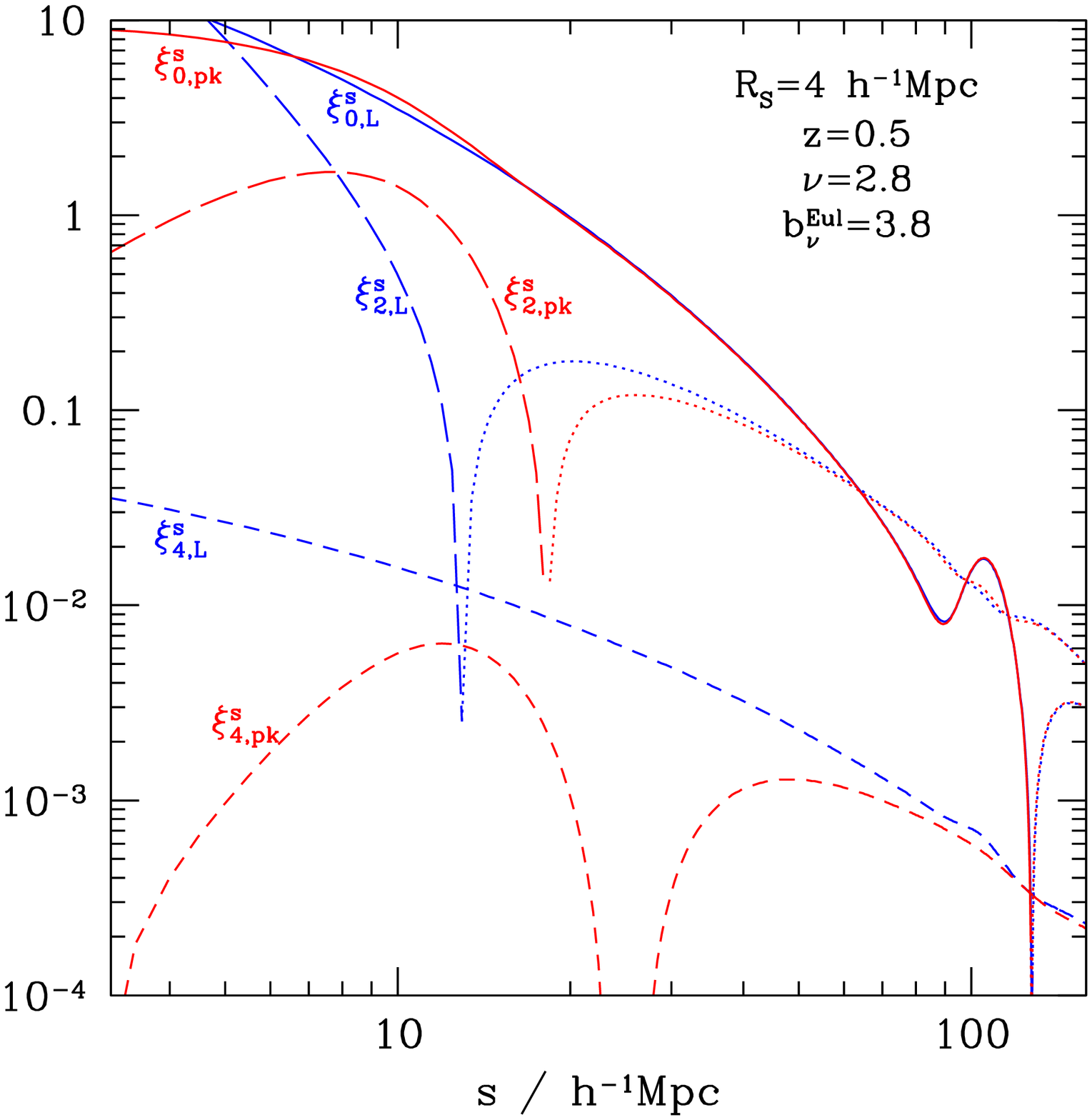}}
\caption{A comparison between the redshift space multipoles of the
correlation function of density maxima and linearly biased tracers.
(Dotted lines denote negative values.)  The peaks were  identified in
the density field when smoothed with a Gaussian filter  of
characteristic scale $R_S=2.5$ (left panel) and $4\hmpc$ (right
panel). This corresponds to a mass scale  $M_S=1.9\times 10^{13}$ and
$7.8\times 10^{13}\hmsun$, respectively. The associated  peak height
and bias parameters quoted in each panel assume a redshift  $z=0.5$.
The linear biased tracers are required to have the same value  of
$b_\nu^{\rm Eul}$.  The peak biasing relation enhances the monopole and
the quadrupole around  the BAO scale relative to that of linearly
biased  tracers, and induces  significant scale dependence in the
hexadecapole at  $s\lesssim 100\hmpc$.}
\label{fig:xis}
\end{figure*}

Here and henceforth, we will assume that $V_{\rm vir}$ is a Gaussian
smoothing kernel.  This implies that the Fourier space multipoles
${\cal P}_\ell^s(k)$ are given by Eqs~(\ref{eq:p0}), (\ref{eq:p2}) and
(\ref{eq:p4}) with
\begin{equation}
\kappa\equiv k \sqrt{\sigma_{\rm vel}^2 f \left(2+f\right)+\sigma_{\rm
vir}^2}\;,
\end{equation}
upon making the replacement
\begin{equation}
\frac{{\cal P}_\ell^s(k)}{\ppk(k)} \to \frac{{\cal P}_\ell^s(k)}{\ppk(k)
e^{-k^2\sigma_{\rm vel}^2}}
\end{equation}
on the left-hand side. We will now illustrate the effect of the
biasing relation Eq.~(\ref{eq:pkbiasing}) on the 2-point correlation
through a comparison between density peaks and linearly biased tracers.

\subsection{Comparison between density peaks and linearly biased tracers}

For linearly biased tracers, $b_\zeta=0$, $\gamma_0=0$ and all the
terms involving $\sigma_0/\sigma_1$ vanish. The pairwise statistics
simplify to 
\begin{align}
  \label{eq:sig12blt}
  v_{12}(r,\mu) &= \frac{-2 b_\nu \xi_1^{(-1/2)}(r)}{1+\xpk(r)}\,
  L_1(\mu)\;, \\
  \sigma_{12}^2(r,\mu) &= 
  \frac{2}{3}\sigma_{-1}^2\left[1 -\frac{\xi_0^{(-1)}}{\sigma_{-1}^2}
    + 2\frac{\xi_2^{(-1)}}{\sigma_{-1}^2}\, L_2(\mu)\right]\;.
\end{align} 
Setting $\beta=f/b_\nu$, we recover the linear theory prediction
of~\cite{Kaiser1987} plus a contribution from the large-scale limit of
$\sigma_{12}^2$ (which underestimates the true effect since we neglect
nonlinear corrections to the velocity dispersion),
\begin{align}
\label{eq:xislt}
 \lefteqn{\xlt^s(s,\mu)=\frac{8}{35}f^2\xi_4^{(0)}L_4(\mu)} \\ &
 \quad -\left[\left(\frac{4}{3}\beta+\frac{4}{7}\beta^2\right)b_\nu^2
 \xi_2^{(0)}-\frac{2}{9}f^2\sigma_{-1}^2 b_\nu^2\xi_2^{(1)}\right]
 L_2(\mu) \nonumber \\ &
 \quad +\left(1+\frac{2}{3}\beta+\frac{1}{5}\beta^2\right)b_\nu^2\xi_0^{(0)}
 -\frac{1}{9}f^2\sigma_{-1}^2b_\nu^2\xi_0^{(1)}\nonumber \;.
\end{align}
Note that Eq.~(\ref{eq:xislt}) implicitly assumes that the peculiar
velocities of linear tracers match locally that of the matter.  To
account for the nonlinear evolution, we will also adopt the
prescription $b_\nu\to b_\nu^{\rm Eul}=1+b_\nu$.

The explicit Legendre decomposition $\xi^s(s,\mu)=\sum \xi_\ell^s(s)
L_\ell(\mu)$  of the redshift space correlation function can be read
off from  Eqs.~(\ref{eq:xispk}) and (\ref{eq:xislt}). For
illustration,  the multipoles $\xi_\ell^s(s,\nu)$ are plotted in the
left and right panel of Fig.\ref{fig:xis} for density maxima
identified at the smoothing scale $R_S=2.5$ and $4\hmpc$,
respectively. These functions are compared to those of linearly biased
tracers with same value of $b_\nu^{\rm Eul}$, namely $b_\nu^{\rm
Eul}=2.0$ and 3.8 respectively. It is important to note that, in the
latter case, the density field is {\it not} smoothed (in practice we
use $R_S=0.1\hmmpc$). Furthermore, we have also neglected the
contribution $\sigma_{\rm vir}$ from virialized motions to the
velocity dispersion and assumed $\sigma^2_{\rm vel}\equiv
\sigma^2_{\rm vpk}/3$ for the peaks and $\sigma_{-1}^2/3$,  with
$\sigma_{-1}^2=(8.11\hmpc)^2$, for the linearly biased tracers.  

\begin{figure*}
\resizebox{0.48\textwidth}{!}{\includegraphics{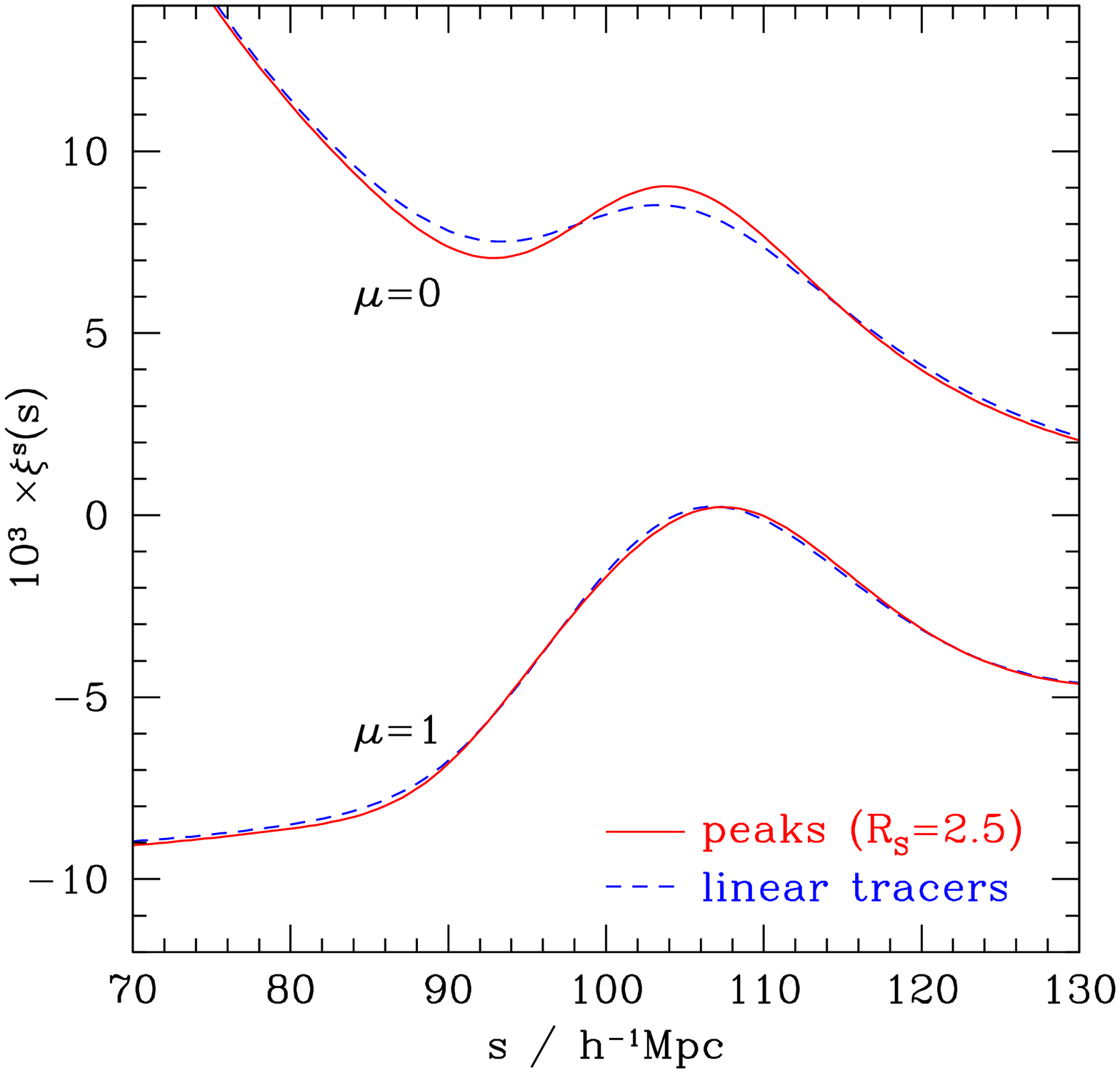}}
\resizebox{0.48\textwidth}{!}{\includegraphics{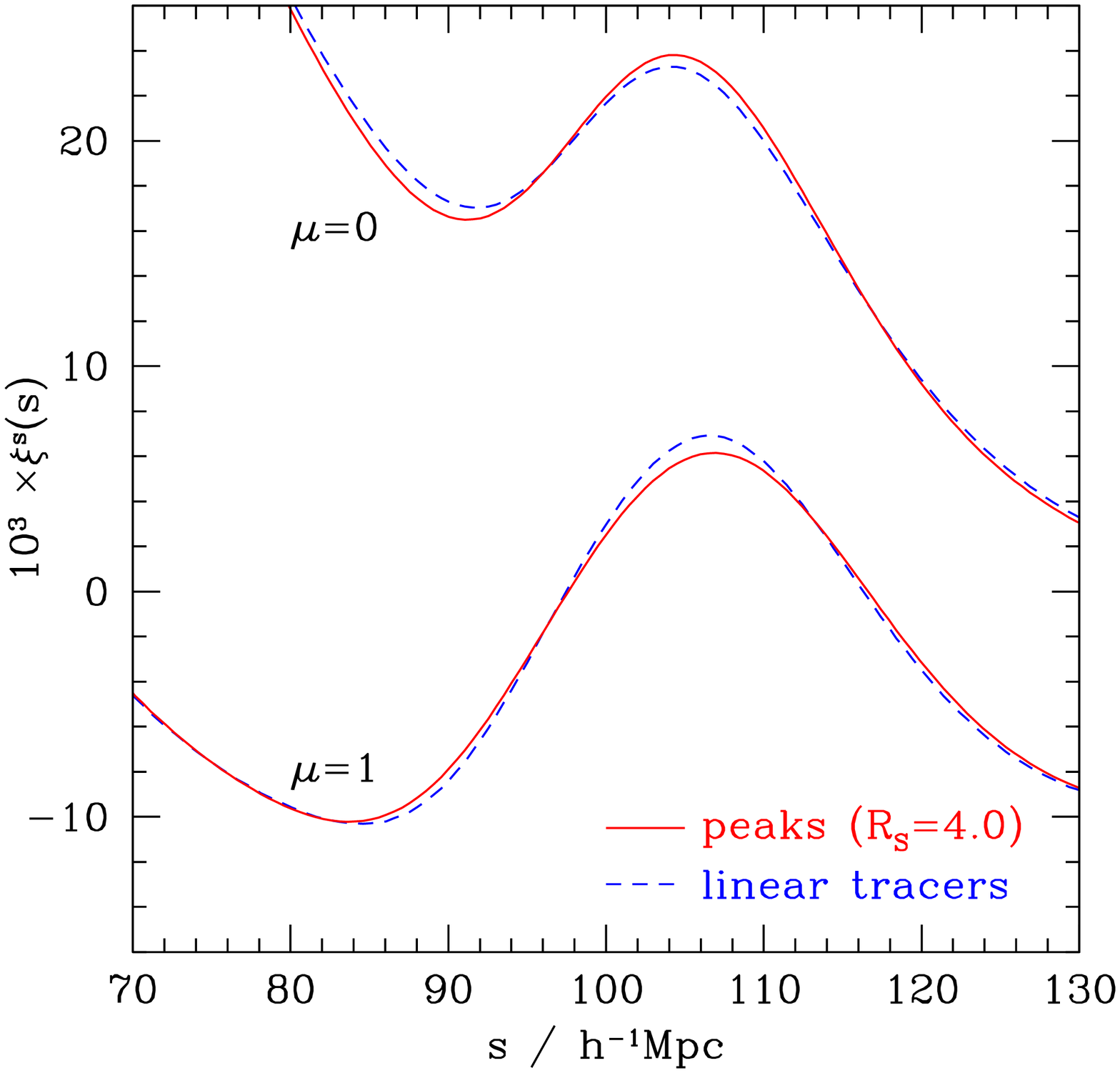}}
\caption{Angular dependence of the redshift space correlation  for the
density maxima (solid curves) and  linearly biased tracers (dashed
curves)  considered in Fig.~\ref{fig:xis}. Results are shown for a
separation vector oriented in the direction parallel ($\mu=1$) and
transverse ($\mu=0$) to the line of sight. Perpendicular to the line
of sight, the  contrast of the  acoustic peak is more pronounced in
the correlation of density peak whereas, in the radial direction, it
is comparable to that of linearly biased tracers.}
\label{fig:los}
\end{figure*}

As recognized in~\cite{Desjacques2008b}, the nonlinear local biasing
relation Eq.~(\ref{eq:pkbiasing}) amplifies the contrast of the  (real
space) baryon acoustic signature of density maxima relative to  that
of linearly biased tracers. A similar enhancement is also observed in
the baryonic acoustic signature of dark matter halos in very large
cosmological simulations \cite{Kimetal2009,Maneraetal2009}. As can be
seen in Fig.\ref{fig:xis}, this amplification is also present in
redshift space. In this case however, both the monopole and the
quadrupole of $\xpk^s$ are affected by the nonlinear peak biasing
across the baryon acoustic oscillation (BAO). At distances $s\sim
100-110\hmpc$, the quadrupole of the redshift space  peak correlation
$\xpk^s$ is indeed more negative, damping thereby the correlation in
the radial direction ($\mu\approx 1$) and increasing it in the
perpendicular direction ($\mu\approx 0$).

This is more clearly seen in Fig.~\ref{fig:los}, which compares the
redshift space correlation of density peaks and linear tracers in the
direction parallel and transverse to the line of sight axis.  Relative
to the baryon acoustic peak of linearly biased tracers, the BAO of
density maxima is enhanced in the direction perpendicular to the line
of sight while somewhat distorted in the radial direction. The
physical origin of this effect presumably is peak-peak exclusion.
Namely, while as discussed in \cite{Desjacques2008b} the spatial bias
of peaks enhances the contrast of the BAO in the real space
correlation, peak-peak exclusion suppresses the infall of peak pairs
onto the (slightly overdense) BAO shell at radius $s\approx
105\hmpc$. In redshift space, this amounts to a reduction of the BAO
contrast along the line of sight. The dispersion term
$\sigma_{12}(\infty)\xi''$  further smoothes the BAO and shifts the
position of the local maximum in that direction, but leaves the baryon
wiggle unchanged in the transverse direction.  The amount of smoothing
depends on the exact value of $\sigma^2_{\rm vel}$.  Another striking
feature of Fig.~\ref{fig:los} is the strong suppression of the
redshift space correlation and the sharpening the acoustic  peak along
the line of sight due to linear coherent infall
~\cite{Kaiser1987,Hamilton1992}.

\begin{figure*}
\resizebox{0.48\textwidth}{!}{\includegraphics{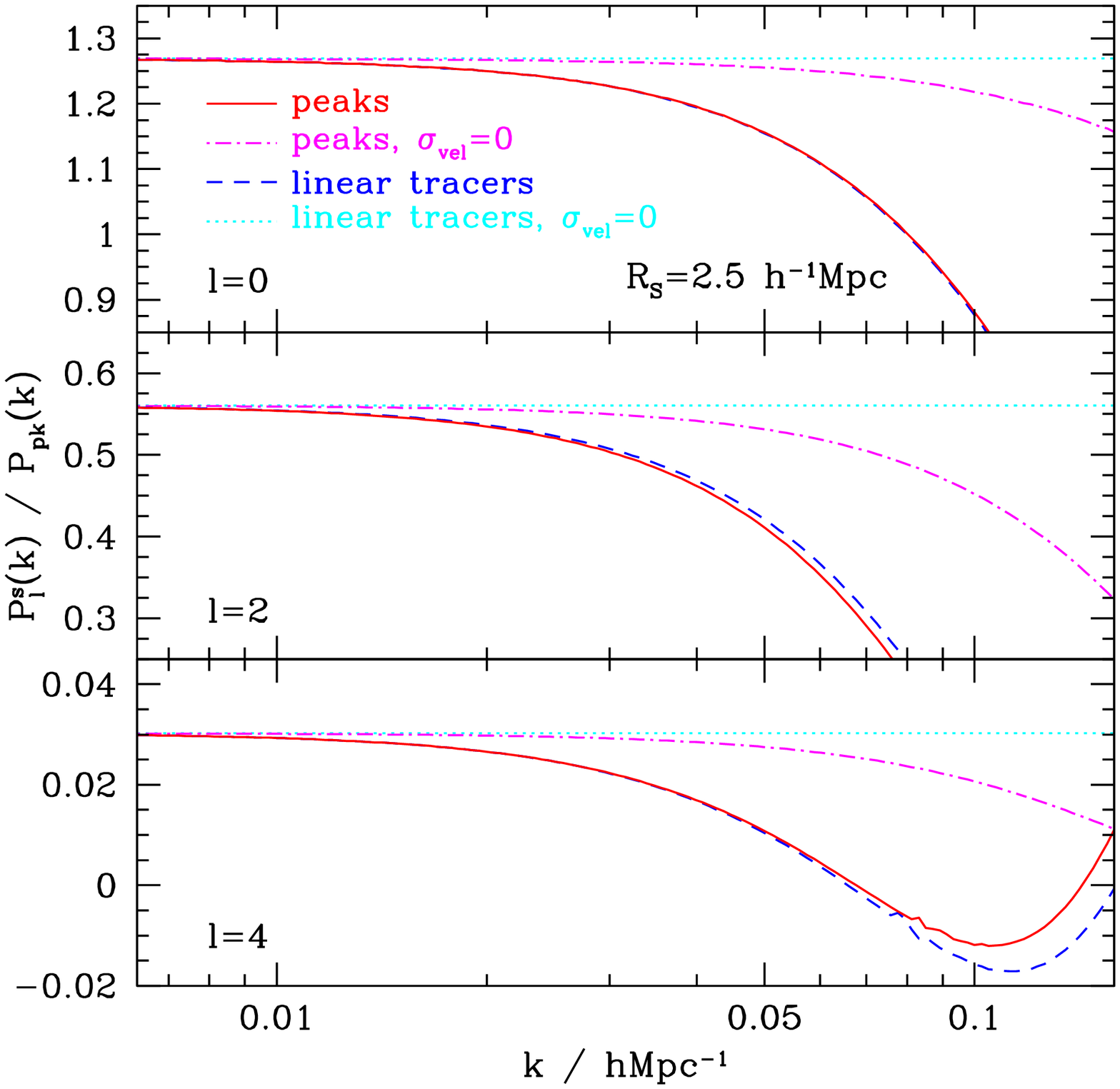}}
\resizebox{0.48\textwidth}{!}{\includegraphics{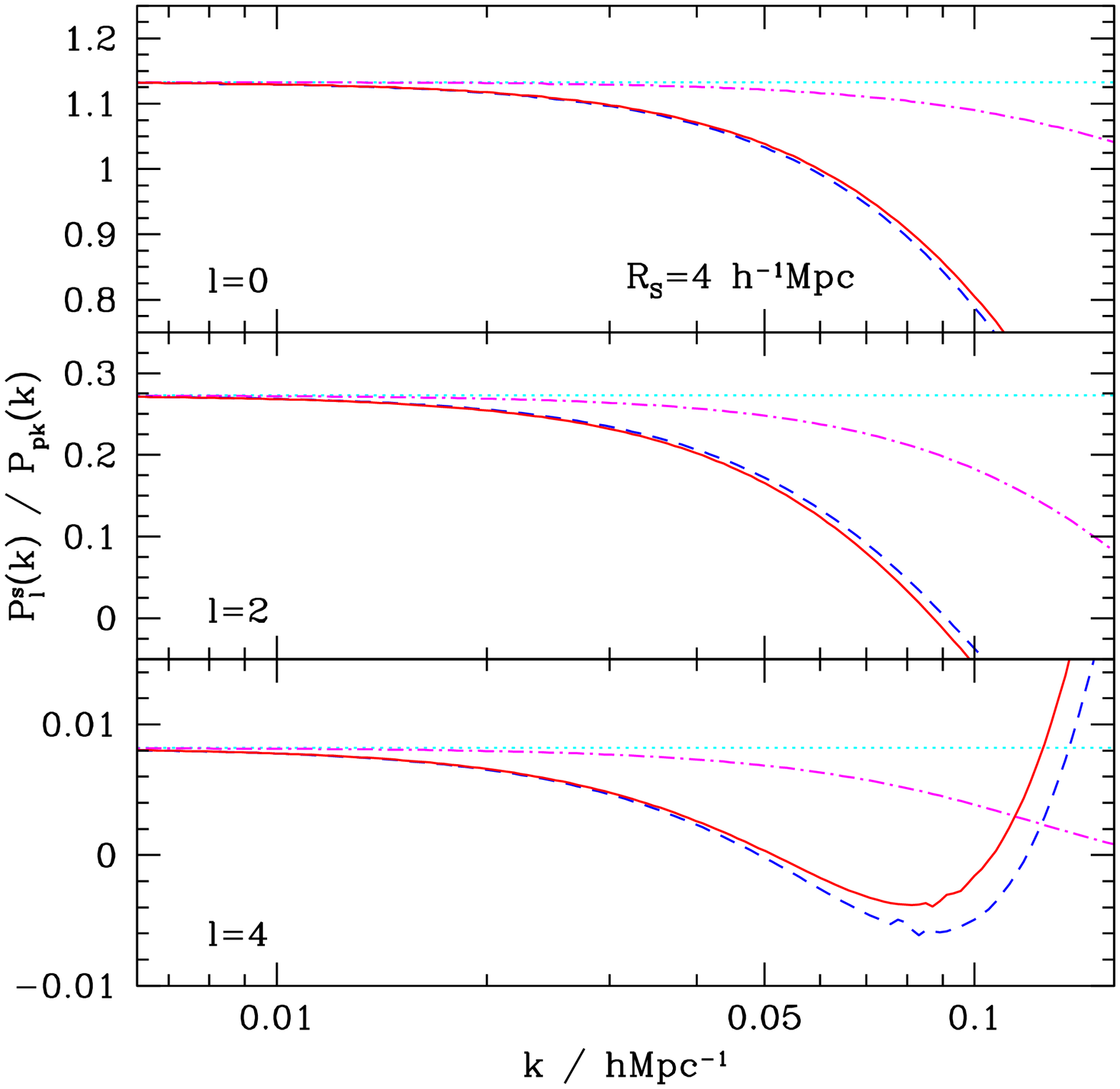}}
\caption{Fourier space multipoles ${\cal P}_\ell^s$ in unit of $P_{\rm
pk}(k)$ as a function  of wavenumber for the density peaks and for the
linearly biased tracers with same value of $b_\nu^{\rm Eul}$.  The
dotted-dashed (peaks) and dotted curves (linear tracers) are the
results without including the velocity damping kernel (assumed to be a
Gaussian, see Sec.~\ref{sub:nonlinear}), while the solid (peaks) and
dashed (linear tracers) curves represent the multipoles when the
Gaussian dispersion is included. For density peaks,  ${\cal
P}_\ell^s(k)$ exhibit a strong $k$-dependence even upon removal  of
the damping term.}
\label{fig:pkl}
\end{figure*}

On scales less than the BAO ring, the contribution of the pairwise
velocity dispersion increases with decreasing separation until it
reverses the sign of the quadrupole at separation $\sim 10-20\hmpc$
and stretches structures along the line of
sight~\cite{Scoccimarro2004}. Although the velocity dispersion of the
density peaks is smaller than that of the linear biased tracers,
peak-peak exclusion makes the effect stronger. On those scales, the
contribution of the term $2b_\nu^{\rm Eul}
b_\zeta\xi_0^{(1)}+b_\zeta^2\xi_0^{(2)}$ becomes comparable to
$(b_\nu^{\rm Eul})^2\xi_0^{(0)}$ and steepens the profile of the
angle-averaged correlation $\xi_0^s$. As a result, the monopole for
the density peaks can be larger by a few tens of per cent at
separation $s\lesssim 10\hmpc$ relative to that of linearly biased
tracers. Note that small-scale halo exclusion is not properly
accounted for in our treatment since we consider $\xpk$ at first order
only. Nevertheless, we expect that, while in real space peak-peak
exclusion leads to a deficit of pairs at distance $s\lesssim R_S$, in
redshift space the suppression may be weaker because peaks tend to
move toward each other.

Fig.~\ref{fig:pkl} displays the Fourier space multipoles ${\cal
P}_\ell(k)$ in unit of $\ppk(k)$  for the peaks and linear tracers
considered above. To emphasize the importance of the exponential
damping, results are shown with and without the smearing caused by
quasi-linear and virialized motions. While for the linearly biased
tracers the distortion parameter ${\cal B}(k)=f/b_\nu^{\rm Eul}$ is a
constant, for peaks ${\cal B}(k)$ is $k$-dependent and, therefore,
induces a scale dependence in the multipoles even when the pairwise
velocity dispersion is negligible. In this limit ($\sigma_{\rm
vel}=0$), for the linearly biased tracers the ratios ${\cal
P}_\ell(k)/\ppk(k)$ are constant (as in the original Kaiser formula),
whereas for peaks  they decay rapidly to reach  $1+2{\cal
B}(\infty)/3+{\cal B}^2(\infty)/5$, $4{\cal B}(\infty)/3+4{\cal
B}^2(\infty)/7$ and  $8{\cal B}^2(\infty)/35$ when $\ell=0$, 2 and 4,
respectively. Here, ${\cal B}(\infty)=-(\nu/\sigma_0-b_\nu^{\rm
Eul})^{-1}$ is the value of ${\cal B}(k)$ in the limit $k\to\infty$
(see Eq.~\ref{eq:blim}).  The difference between peaks and linear
tracers is largest in the hexadecapole and increases with mass
scale. For instance, the fractional deviation is 5 per cent at
wavenumber $k\approx 0.037\hmmpc$  and $\approx 0.027\hmmpc$ for the
peaks identified at filtering scale $R_S=2.5$ and 4$\hmpc$,
respectively.

When the Gaussian damping term, Eq.~(\ref{eq:damping}), is included,
the behaviour of the Fourier space multipoles of density peaks (solid
curves) and linear tracers (dashed curves) becomes similar at
small-scale: they damp to zero like the coefficients $A_\ell(\kappa)$
defined in Eq.~(\ref{eq:acoeff}). Still, significant deviations
persist on scale $k\gtrsim 0.01\hmmpc$ due to the $k$-dependence of
${\cal B}(k)$ and unequal velocity dispersions.

\section{Cosmological implications}
\label{sec:cosmo}

\subsection{Estimating the growth rate $f$}

Following \cite{Kaiser1987,PeacockDodds1994,Scoccimarro2004}, the
redshift space power spectrum of density peaks can also be written as
\begin{equation}
\ppk^s(k)=\Bigl[\ppk(k)+2\mu^2\ppt(k)+\mu^4\ptt(k)\Bigr] F(k,\mu^2)
\end{equation}
where $\ppt(k)$ and $\ptt(k)$ are the peak-velocity and
velocity-velocity power spectra. Here, $\ppk$, $\ppt$ and $\ptt$ are
{\it linear} spectra and $F(k,\mu^2)=V_{\rm ql}(k,\mu^2) V_{\rm
vir}(k,\mu^2)$ describes both the quasi-linear damping and the
smearing from the small-scale velocity dispersion. As noted in
\cite{SongPercival2008,PercivalWhite2009,Whiteetal2009,Tayloretal2001},
$\ptt(k)$ is independent of the spatial bias and directly measures the
matter velocity power spectrum provided there is no velocity
bias. Owing to the angular dependence, a measurement of the velocity
power spectrum furnishes an estimate of the linear growth rate $f
\sigma_8\propto dD/d\ln a$ that is not affected by the spatial bias.

Although the hexadecapole does not depend upon the spatial bias, it
may be noisier than the monopole and dipole, so this has motivated the
search for other combinations of $P_0$ and $P_2$ which may be more
robust \cite{Hamilton1992}.  
Reference \cite{PercivalWhite2009} showed that ${\cal P}_0^s$
and ${\cal P}_2^s$ can be used to derive an estimate of the velocity
power spectrum $\ptt(k)$ when the density and velocity fields are
perfectly correlated, namely, when the cross-correlation coefficient
\begin{equation}
r_\theta^2(k)=\frac{\ppt^2(k)}{\ppk(k)\ptt(k)}
\end{equation}
is unity. For example, when smoothing is ignored, then
\begin{equation}
  \hat{P} \equiv \frac{245}{48} P_0^s\,\left(1 + \frac{P_{20}^s}{7} -
  \sqrt{1 + \frac{2P_{20}^s}{7}  -
  \frac{\left(P_{20}^s\right)^2}{5}}\right)
\end{equation}
\cite{PercivalWhite2009} is proportional to $f^2\, P_\delta(k)$ when
velocities are unbiased. Here, $P_{20}^s\equiv P_2^s/P_0^s$. For peaks
$r_\theta^2(k)\equiv 1$ indeed holds at the lowest order, even though
the linear spatial and velocity bias $b_{\rm pk}(k)$ and $b_{\rm
vel}(k)$ are scale dependent. However, the velocity power spectrum now
is $\ptt(k)=f^2 b_{\rm vel}^2(k) P_\delta(k)$, so there is an extra
$k$-dependence associated with the estimator $\hat{P}$.  Since this
could be interpreted erroneously as a signature of modified dark
energy or gravity, any scale dependent velocity bias (a scale
independent bias may also be present if the tracers do not move with
the matter) will limit the information that can be recovered about the
growth factor
\cite{SongPercival2008,PercivalWhite2009,Whiteetal2009}. For peaks,
the velocity bias is $b_{\rm vel}(k)\leq 1$, and it converges towards
unity (i.e. unbiased velocities) in the limit $k\to 0$. At the first
order, the deviation from unity is controlled by $\sigma_0/\sigma_1$
(Eq.~\ref{eq:vpkbiask}) so that, at fixed wavenumber, $b_{\rm vel}(k)$
decreases with increasing mass scale (see Fig.~\ref{fig:biases}. For
$M_S=1.9$ and $7.8\times 10^{13}\hmsun$ considered here, $\ptt(k)$ is
suppressed by $\approx$ 5 and 9 per cent at wavenumber $k=0.05\hmmpc$,
respectively.  The predicted $k$-dependence is smaller than current
constraints on the growth rate
\cite{NesserisPerivolaropoulos2008,SongPercival2008}.  Furthermore,
numerical simulations to date show that the power spectrum of dark
matter halo velocities is consistent with $f^2 P_\delta(k)$ within 10
per cent at wavenumber $k\lesssim 0.1\hmmpc$
\cite{PercivalWhite2009}. Nevertheless, since forthcoming large-scale
galaxy surveys will dramatically improve constraints on the growth
factor (down to the percent level), it is interesting to assess the
extent to which a $k$-dependent bias would degrade the constraint on
the growth rate.

\subsection{Error forecast with a $k$-dependent velocity bias}

To this purpose, we use the Fisher based formalism developed in
\cite{Whiteetal2009}. For Gaussian random fields, the Fisher matrix
for a set of parameters $\{p_i\}$ is
\cite{VogeleySzalay1996,Tegmarketal1998}
\begin{equation}
  F_{ij}=\frac{1}{2}\int\!\!\frac{d^3k}{(2\pi)^3}\left(\frac{\partial\ln
  P} {\partial\ln p_i}\right)\left(\frac{\partial\ln P}{\partial\ln
  p_j}\right)   V_{\rm eff}(\vk)\;,
\end{equation}
where $P$ is the power spectrum and individual wavemode contributions
are weighted by the effective volume \cite{Feldmanetal1994}
\begin{equation}
V_{\rm eff}(\vk)=V \left(\frac{\bar{n}P}{1+\bar{n}P}\right)^2
\end{equation}
which depends upon the surveyed volume $V$ and the number density
$\bar{n}$ of the tracers (assumed homogeneously distributed).  
To illustrate, we assume the linear, plane-parallel approximation 
and consider the model
\begin{equation}
P^s(k,\mu)=\Bigl[b_{\rm pk}(k)+f b_{\rm vel}(k)\mu^2\Bigr]^2
P_\delta(k)\;,
\end{equation} 
where
 $b_{\rm pk}(k)\equiv b_\nu+b_\zeta k^2$ (we drop the superscript
{\small Eul} for brevity), and
 $b_{\rm vel}(k)\equiv 1- R_{\rm vel}^2 k^2$ (for some $R_{\rm vel}$) 
are motivated by the functional form of the spatial and velocity bias 
of density peaks (c.f., Section~\ref{sec:bias} and~\ref{sec:velb}).  

In what follows, we fix the shape and amplitude of the matter power 
spectrum (i.e. the fractional error on $f\sigma_8$ is equal to that 
on $f$) and  consider the four-parameter set
 $\left\{b_\nu,b_\zeta,R_{\rm vel},f\right\}$. 
Our fiducial model has $\left(b_\nu,b_\zeta,R_{\rm vel},f\right)=\left(
1,16,3,0.46\right)$. The values of $b_\zeta$ and $R_{\rm vel}$ closely
correspond to those of density peaks identified at the mass scale
$1.9\times 10^{13}\hmsun$.  Derivatives of the logarithm of the power
with respect to the parameters are computed easily:
\begin{gather}
 \frac{\partial\ln P}{\partial b_\nu}=
 \frac{2} {\left(b_{\rm pk} + fb_{\rm vel}\mu^2\right)}, \quad
 \frac{\partial\ln P}{\partial b_\zeta}= 
 \frac{2k^2} {\left(b_{\rm pk}+fb_{\rm vel}\mu^2\right)}, \nonumber\\
 \frac{\partial\ln P}{\partial R_{\rm vel}}= 
 \frac{-4f R_{\rm vel}\mu^2 k^2} {\left(b_{\rm pk}
   + f b_{\rm vel}\mu^2\right)} \nonumber \\ 
 \frac{\partial\ln P}{\partial f}=
 \frac{2 b_{\rm vel}\mu^2} {\left(b_{\rm pk}+fb_{\rm vel}\mu^2\right)}\;.
\end{gather}
We integrate over wavenumbers from $k_{\rm min}\sim \pi/V^{1/3}$,
where $V$ is the volume of the survey, up to a maximum wavenumber
$k=0.1\hmmpc$, above which nonlinear effects are expected to become
important \cite{Whiteetal2009}.

In order to illustrate the effect of  including a $k$-dependent
velocity bias into the analysis, we initially set $b_{\rm vel}\equiv
1$ (i.e. ignore $R_{\rm vel}$) and compute the  Fisher matrix for
$b_\nu$, $b_\zeta$ and $f$ solely. For a survey of  volume
$V=10\hgggpc$ at redshift $z=0$, we find a fractional  marginalized
error of  $\delta f/f=$1.6\% in the limit $\bar{n}P\gg 1$  of
negligible shot noise (In practice, a suitable weighting of galaxies
may help approaching this limit \cite{Seljaketal2009}).
For a number density $\bar{n}=5\times 10^{-4}$ and $10^{-4}\hhhmpc$,
the constraint weakens to 1.9\% and 2.9\%, respectively.  (These
values are consistent with those of \cite{Whiteetal2009}.)
Unsurprisingly, $b_\nu$ and $b_\zeta$ are strongly anti-correlated
(the correlation coefficient  is $r\approx -0.8$) because an increase
in $b_\nu$ can be mostly compensated by a decrease in
$b_\zeta$. However, while the correlation between $b_\nu$ and $f$ is
moderate ($r\lesssim -0.5$), $b_\zeta$ and $f$ are weakly degenerate
($r\lesssim -0.05$). In other words, including a $k$-dependent bias
component $b_\zeta k^2$ has little effect on the uncertainty on $f$.
Extending $k_{\rm max}$ beyond $0.1\hmmpc$ (where the shot noise
becomes again important) can reduce the uncertainty on $f$ (because
the fractional error scales as  $k_{\rm max}^{-3/2}$), but this is at
the price of having to model the  smearing due to quasi-linear motions
and small-scale velocities.

Introducing the parameter $R_{\rm vel}$ substantially increases the
uncertainty on $f$. For the volume $V$ and the average number
densities $\bar{n}$ considered above, the fractional marginalized
uncertainty on the growth rate becomes $\delta f/f$=4\%, 4.4\% and
6\%, respectively.  This can be traced to the strong correlation
($r\approx 0.9$) between $R_{\rm vel}$ and $f$. The error degradation
reflects the fact that we are adding more freedom to the model. It
does not depend upon the exact  value of  $R_{\rm vel}$.

Are the constraints on $f$ obtained using the multi-tracer method
proposed in \cite{Seljak2009} affected in a similar way~?
Reference \cite{McDonaldSeljak2008} pointed out that several
populations of differently biased  tracers can achieve a much better
determination of the growth rate than a single sample of objects.
When power spectra are measured,  calculating the Fisher matrix for
multiple tracers requires  summing over the distinct components of the
inverse covariance matrix $C_{AB}^{-1}$, where $A$, $B$ label a
different pair of tracer populations,
\begin{equation}
  F_{ij}=V\sum_{A,B} \int\!\!\frac{d^3
  k}{(2\pi)^3}\left(\frac{\partial P_A} {\partial p_i}\right)
  C_{AB}^{-1} \left(\frac{\partial P_B}{\partial p_j} \right)\;.
\end{equation}
The calculation of the covariance matrix is straightforward if one
assumes that the noise term can be treated as an uncorrelated normal
variate  \cite{Whiteetal2009,BurkeyTaylor2004}. For completeness, the 
diagonal and off-diagonal components of the covariance matrix are 
\cite{Whiteetal2009}
\begin{equation}
\la C_{aaaa}\ra = 2 P_{aa}^2 N_a^2, \quad \la
 C_{abab}\ra = P_{ab}^2+P_{aa}P_{bb}N_a N_b
\end{equation}
and
\begin{gather}
\la C_{abcd}\ra=2 P_{ab} P_{cd},\quad  \la C_{aabc}\ra=2 P_{ab}
P_{ac} \nonumber \\ 
\la C_{abac}\ra=P_{ab}P_{ac}+P_{aa}P_{bc}N_a \\ 
\la C_{aaab}\ra=P_{ab}P_{aa}N_a,\quad \la C_{aabb}\ra=2 P_{ab}^2 
\nonumber \;.
\end{gather}
where $P_{ij}$ is an auto- or cross-power spectrum and
 $N_a\equiv \left[1+1/(\bar{n}P_{aa})\right]$. 
We will consider the simplest case of two types of tracers, since 
the gains saturate rapidly as the number of samples increase 
\cite{Whiteetal2009}.  In this case there are 3 distinct measured 
power spectra
\begin{equation}
 P_{ab}= \left(b_{\rm pk}^{(a)}+f b_{\rm vel}^{(a)}\,\mu^2\right)\,
 \left(b_{\rm pk}^{(b)}+f b_{\rm vel}^{(b)}\,\mu^2\right)\, P_\delta(k)
\end{equation}
where $a,b=1,2$ and the bias factors are
 $b_{\rm pk}^{(a)} = b_\nu^{(a)}+b_\zeta^{(a)}k^2$,  
 $b_{\rm vel}^{(a)} = 1 - (R_{\rm vel}^{(a)})^2 k^2$. 
The power spectra are thus described by 7 parameters, and their 
derivatives are
\begin{align}
\frac{\partial P_{ab}}{\partial b_\nu^{(c)}} &= \biggl[\left(b_{\rm
pk}^{(b)}+f b_{\rm vel}^{(b)}\mu^2\right)\delta_{ac}^K
\biggr. \\ & \qquad +\biggl.\left(b_{\rm pk}^{(a)} +f b_{\rm
vel}^{(a)}\mu^2\right)\delta_{bc}^K\biggr] P_\delta(k) \nonumber \\
\frac{\partial P_{ab}}{\partial b_\zeta^{(c)}} &= k^2
\biggl[\left(b_{\rm pk}^{(b)}+f b_{\rm
vel}^{(b)}\,\mu^2\right)\delta_{ac}^K \biggr. \\ & \qquad +
\biggl.\left(b_{\rm pk}^{(a)}+f b_{\rm vel}^{(a)}\,\mu^2\right)
\delta_{bc}^K\biggr] P_\delta(k) \nonumber \\ \frac{\partial
P_{ab}}{\partial R_{\rm vel}^{(c)}} &= -2f k^2\mu^2
\biggl[\left(b_{\rm pk}^{(b)}+f b_{\rm vel}^{(b)}\,\mu^2\right) R_{\rm
vel}^{(a)} \delta_{ac}^K\biggr.\\ & \qquad + \biggl.\left(b_{\rm
pk}^{(a)}+f b_{\rm vel}^{(a)}\,\mu^2\right) R_{\rm vel}^{(b)}
\delta_{bc}^K\biggr] P_\delta(k) \nonumber \\ \frac{\partial
P_{ab}}{\partial f} &= \mu^2 \biggl[\left(b_{\rm pk}^{(b)} +f b_{\rm
vel}^{(b)}\,\mu^2\right) b_{\rm vel}^{(a)}\biggr. \\  &
\qquad + \biggl.\left(b_{\rm pk}^{(a)}+f b_{\rm vel}^{(a)}\,
\mu^2\right)b_{\rm vel}^{(b)}\biggr] P_\delta(k) \nonumber \;.
\end{align}
Here, $\delta_{ab}^K$ is the Kronecker delta.

\begin{figure}
\resizebox{0.48\textwidth}{!}{\includegraphics{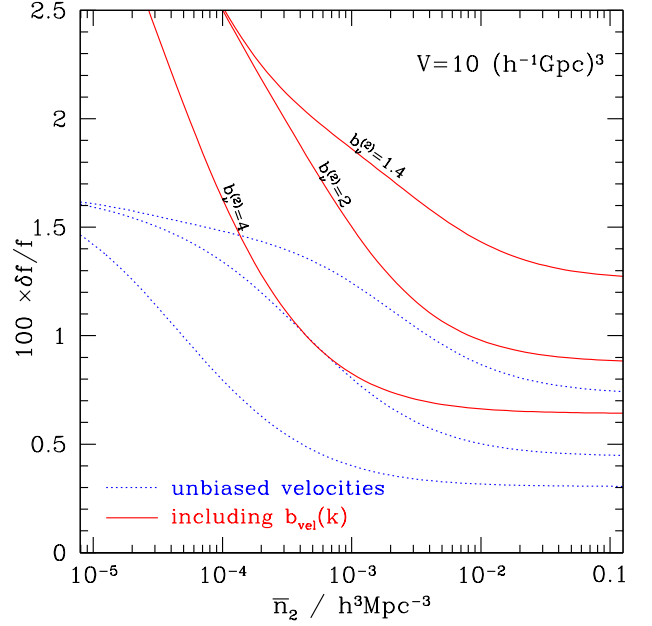}}
\caption{The fractional marginalized error $\delta f/f$ for a survey
volume $V=10\hgggpc$ at $z=0$ obtained with two tracer populations~: a
high density, unbiased sample with $\bar{n}_1=10^{-2}\hhhmpc$ and
$b_\nu^{(1)}=1$ and a second population with varying number density
$\bar{n}_2$ and bias $b_\nu^{(2)}$. We choose $b_\nu^{(2)}=1.4$, 2 and
4 (curves from top to bottom). The constraints are shown as a function
of $\bar{n}_2$ assuming  $b_{\rm vel}^{(1)}=b_{\rm vel}^{(2)}\equiv 1$
(dotted curves) and a $k$-dependent velocity bias with $R_{\rm
vel}^{(1)}=R_{\rm vel}^{(2)}=3\hmpc$ (solid curves).}
\label{fig:margerr}
\end{figure}

Fig.~\ref{fig:margerr} shows the fractional marginalized error on $f$
obtained by combining two different biased sample of the same survey
volume. The constraints are shown as a function of the abundance of
the second tracers with and without including a $k$-dependent velocity
bias (solid and dotted curves, respectively). We set
$b_\nu^{(2)}=1.4,$ 2 and 4 to facilitate the comparison with Fig.3 of
\cite{Whiteetal2009}. Although we have assumed the fiducial values
$R_{\rm vel}^{(1)} = R_{\rm vel}^{(2)} = 3\hmpc$ for simplicity, we may
expect from the analysis done in the previous Section that 
$R_{\rm vel}$ has some mass or bias dependence.  
E.g., $R_{\rm vel}^{(2)}>R_{\rm vel}^{(1)}$ when 
$b_\nu^{(2)}\gg b_\nu^{(1)}$.  
The marginalized error on $f$ is, however, weakly dependent on the
fiducial value of $R_{\rm vel}^{(a)}$.  Note the considerable 
improvement in the constraint on $f$ 
\cite[in agreement with the findings of][]{McDonaldSeljak2008,Whiteetal2009}. 
The smallest error is achieved with a large number density $\bar{n}_2$ 
and large relative bias $b_\nu^{(2)}/b_\nu^{(1)}$.  (We have used 
values of $b_\nu^{(2)}$ to simplify comparison with \cite{Whiteetal2009}.) 
However, including a $k$-dependent velocity bias degrades the
uncertainty on the growth rate roughly by a factor of two when
$\bar{n}_2\gtrsim 10^{-2}\hhhmpc$, like in the single tracer case. The
error degradation becomes increasingly severe as one goes to lower
number densities.  
 
Although these constraints are only indicative (We have ignored the
influence of cosmological parameters on the constraint
\cite{Striletal2009}), our analysis demonstrates that allowing for a
$k$-dependent velocity bias (with the specific functional form
predicted by the peak model) has a large impact on the determination
of the growth factor and, therefore, may possibly hamper our ability
to distinguish between different dark energy or gravity scenarios
\cite{JainZhang2008,SongKoyama2008}.  Therefore, despite the lack  of
current evidence for a $k$-dependent velocity bias
\cite{PercivalWhite2009}, it seems prudent to study this possibility
further with large cosmological simulations.  We hope the peak model
can serve as a useful baseline with which to compare the simulations.

\section{Stochasticity}
\label{sec:stoch}

\subsection{Cross-correlation coefficient}

The biasing eq.~(\ref{eq:pkbiasing}) derived from the large-scale
properties of peak correlation functions is a mean bias relation that
does not contain any information about stochasticity.  Therefore, it
is unsurprisingly deterministic like the local bias model considered
by \cite{FryGaztanaga1993}, the main difference residing in the fact
that peak biasing involves derivatives of the density field. Still,
because of the discrete nature of density peaks, one can expect that
the peak overdensity $\delta\npk$ at location $\vx$ generally be a
random function of the underlying matter density (and its derivatives)
in some neighbourhood of that point. We note that  stochastic models of
the form $\delta\npk(\vx)=X[\delta_S(\vx)]$  have been studied in
\cite{ScherrerWeinberg1998,Pen1998,DekelLahav1999} for instance.

Computing the probability of $X$ given $\delta$ etc. is beyond the
scope of this paper (because it requires the full hierarchy of
correlation functions). Still, it is instructive to compute the
cross-correlation coefficient to gain further understanding of the
peak biasing model.  The cross-correlation coefficient is defined as
\begin{equation}
 r_c^2(k)   = \frac{\ppd^2(k)}{\ppk(k)P_\delta(k)},~~ r_\xi^2(r) =
 \frac{\xpd^2(r)}{\xpk(r)\xi_\delta(r)}
\end{equation}
in Fourier and configuration space, respectively. Ignoring the damping
term, we find $r_c(k)=1$ for peaks even though the ratio
$\ppd/P_\delta$ depends on $k$.  Thus, a $k$-dependent bias at the
linear order does  not yield stochasticity in Fourier space.  On the
other hand,
\begin{equation}
r_\xi^2(r)= \frac{\left(b_\nu +
 b_\zeta\xi_0^{(1)}/\xi_0^{(0)}\right)^2} {b_\nu^2 + 2 b_\nu b_\zeta
 \xi_0^{(1)}/\xi_0^{(0)}  + b_\zeta^2 \xi_0^{(2)}/\xi_0^{(0)}}\;.
\label{rxi}
\end{equation}
Therefore, although the bias is deterministic in Fourier space, it is
generally stochastic and scale dependent in configuration
space. However, when $\nu\gg 1$ then $b_\zeta\to 0$, so $r_\xi\to
1$. Namely, in the high peak limit, the bias becomes linear and
deterministic in both Fourier and configuration space.

\begin{figure}
\resizebox{0.48\textwidth}{!}{\includegraphics{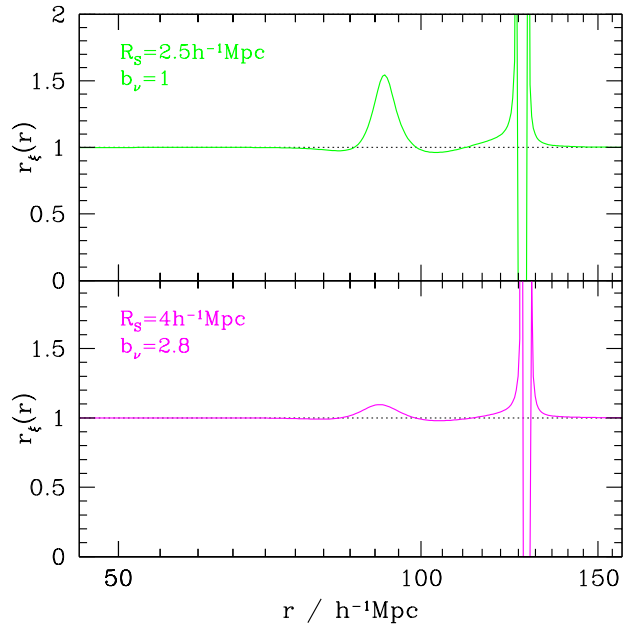}}
\caption{The cross-correlation coefficient $r_\xi(r)$ for density
peaks identified at the smoothing scale $R_S=2.5$ and $4\hmpc$ (upper
and lower panel, respectively). There is significant stochasticity
only at the zero-crossings of the auto and cross-correlation functions
and across the BAO, where the derivatives of the density correlation
$\xi_0^{(1)}$ and $\xi_0^{(2)}$ are not negligible
\cite{Desjacques2008b}.}
\label{fig:rcross}
\end{figure}

The real space cross-correlation  coefficient is shown in
Fig.~\ref{fig:rcross} as a function of comoving separation for the
peaks identified at the smoothing radius $R_S=2.5$ and $4\hmpc$.
$r_\xi$ is very close to unity at all separations larger than a few
smoothing radii, except around the baryonic bump and the zero crossing
of the correlation function where it can be noticeably larger than
unity. These findings seem to run contrary to the common knowledge
that $|r_\xi|\leq 1$. However, at separation $r\gtrsim 120\hmpc$, the
fact that the zero-crossings  of $\xi_0^{(n)}$ do not generally
coincide unavoidably implies  $|r_\xi|>1$, at least over some range of
scales. Furthermore, at distance $r\sim 90\hmpc$, the large values of
$r_\xi$ are most plausibly  traced to the baryon acoustic feature,
which induces large oscillations in $\xi_0^{(1)}$ and $\xi_0^{(2)}$
across the BAO scale $\approx 105\hmpc$ (see Fig.1 of
\cite{Desjacques2008b}).   At the level of a bias relation
$\delta\npk=X[\delta_S,\nabla^2\delta_S,\cdots]$, this suggests that
the scatter is strongly sensitive to $\nabla^2\delta_S$.

\begin{figure}
\resizebox{0.48\textwidth}{!}{\includegraphics{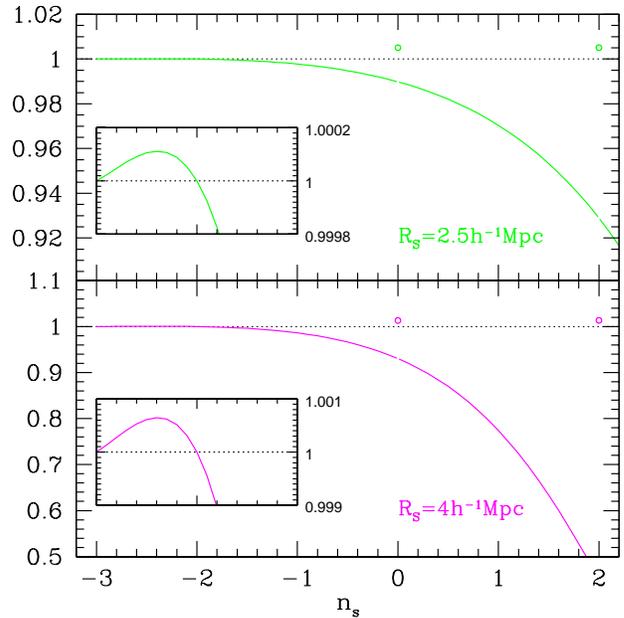}}
\caption{Cross-correlation coefficient $r_\xi(r)$ for powerlaw  power
spectra $P_\delta(k)\propto k^{n_s}$ as a function of the spectral
index $n_s$. Results are shown at a single separation $r=100\hmpc$,
for density peaks identified at the smoothing scale $R_S=2.5$ (top)
and $4\hmpc$  (bottom). The insert is an enlarged view of $r_\xi$ in
the range $-3<n_s<-1$. Notice the discontinuities at $n_s=0$ and 2, at
which the cross-correlation  coefficient is slightly larger than
unity.}
\label{fig:rcplaw}
\end{figure}

Figure \ref{fig:rcplaw} explores the behaviour of the
cross-correlation  coefficient when the underlying power spectrum  is
a featureless power law spectrum, $P_\delta(k)\propto k^{n_s}$.
Results are presented as a function of the spectral index $n_s$ for a
single value of the separation, $r=100\hmpc$.  At fixed value of
$n_s$, the stochasticity is unsurprisingly larger for the relatively
sparser peaks identified at scale $R_S=4\hmpc$. Most importantly, the
amount of stochasticity depends sensitively upon the shape of the
matter power spectrum.  Overall, $r_\xi$ decreases with larger values
of the powerlaw exponent $n_s$ because the stochasticity rises as the
relative  amount of small-scale power increases. As can also be seen,
$r_\xi$ is slightly larger than  unity in the range $-3<n_s<-2$ and at
the points of discontinuity  $n_s=0,2$ (which are marked as empty
symbols). Although the effect is admittedly small and localized in
$n_s$, this demonstrates that the cross-correlation coefficient can
exceed unity also when the power spectrum is scale-free.  In Appendix
\ref{sec:app2}, we investigate the discontinuities in more detail and
provide quantitative estimates of the large-scale behaviour of the
cross-correlation  coefficient for a few values of $n_s$.

\subsection{Evolution of stochastic bias}

As discussed in Sec.\ref{sub:nonlinear}, gravitational evolution maps
a scale independent, deterministic linear bias factor in the initial
conditions onto a similar quantity  in the evolved distribution
\cite{MoWhite1996}.
The scaling $b_\nu^{\rm Eul}=1+b_\nu$ also works for a $k$-dependent
deterministic bias. More precisely,
\begin{equation}
 b^{\rm Eul}_{\rm pk}(k,z) - 1 = b_{\rm pk}(k,z)  = \frac{b^{\rm
  Eul}_{\rm pk}(k,z_0) - 1}{D(z)/D(z_0)}  = \frac{b_{\rm
  pk}(k,z_0)}{D(z)/D(z_0)}
 \label{bevolution}
\end{equation}
where $D(z)$ is the linear theory growth factor \cite{Fry1996,
HuiParfrey2008}.  This is easily understood if one  recognizes that a
peak of height $b_z\delta_z$, where $\delta_z$ is  the linearly
evolved field at $z$, could also have been written as  having height
$b_0\delta_0$ where $\delta_0$ is the field evolved  to $z_0$. The
relation $b_z\delta_z = b_0\delta_0$ implies  $b_0 = b_z\,
(\delta_z/\delta_0) = b_z\,D(z)/D(z_0)$, from which  the above
expression is derived.  Alternatively, notice that  $b_0\propto
1/\sigma_0(z_0) \propto D(z)/D(z_0)/\sigma_0(z)  \propto b_z\,
D(z)/D(z_0)$, so the factor $D(z)/D(z_0)$ is  simply converting from
one choice of fiducial time to another.

In configuration space, it has been argued that for linear  stochastic
bias,
\begin{equation}
 b_\xi^{\rm Eul}(z) r_\xi(z) - 1  = \frac{b_\xi^{\rm
  Eul}(z_0)\,r_\xi(z_0) - 1}{D(z)/D(z_0)}
 \label{brevolution}
\end{equation}
\cite[e.g.][]{TegmarkPeebles1998}.  The corresponding expression  for
the evolution of $b_\xi^{\rm Eul}(r,z)$ itself is
\begin{align}
 b_\xi^{\rm Eul}(z)^2\,D^2(z) &= b_\xi^{\rm Eul}(z_0)^2\,D^2(z_0) 
 + [D(z_0)-D(z)]^2 \nonumber \\ & \quad
 - 2\,[D(z_0)-D(z)]\, D(z_0) \nonumber \\ & \qquad
 \times b_\xi^{\rm Eul}(z_0) r_\xi(z_0) \;.
\end{align}
This is a good model of $b_\xi^{\rm Eul}r_\xi$ for density peaks,
provided we interpret the denominator of Eq.~(\ref{rxi}) as
$b_\xi^2$. Moreover, the numerator of this equation is similar to (the
square of) an Eulerian bias factor minus one:  $([1+b_\nu +
b_\zeta\xi_0^{(1)}/\xi_0^{(0)}] - 1)^2$.  This quantity clearly scales
with the growth factor like its Fourier space analog. Hence, the real
space evolution of the stochastic bias of density  peaks is simple in
spite of the additional scale dependence.  Notice that both $b_{\rm
pk}^{\rm Eul}(k)$ and $b_\xi^{\rm Eul} r_\xi$ tend to unity at late
times (even though they might effectively not reach this limit because
the growth factors freeze out in $\Lambda$CDM-like models). In fact
$b_\xi$ does as well, but this is not as easy to see from our
expressions.

\subsection{Connection to previous work}

We mentioned earlier that peak bias and its evolution have been
studied  in simulations by \cite{MoJingWhite1997,Taruya2001}.  These
authors found that  the peak background split argument
(Eq.~\ref{eq:bpbs}) provides a  good description of the large scale
bias of the peaks extracted from their simulations.   They also found
that Eq.~(\ref{bevolution}) is in reasonable agreement with the
evolution of this large scale bias.   However, on smaller scales, the
real space bias was found to be  scale dependent and
stochastic~\cite{Taruya2001}, two features which a peak-background
split based analysis does not model.  Nevertheless,
Eqs~(\ref{bevolution}) and~(\ref{brevolution}) were found to  provide
a good description of the evolution.  The above analysis shows why.
It would be interesting to see if  our approach correctly predicts the
scale dependence of the bias. We defer this issue to a future work.

\section{Discussion and conclusions}

We have presented an extensive analysis of the Gaussian peak model.
Density peaks are biased tracers of the underlying matter density
field -- on large scales this bias is scale independent -- and we
studied the limit in which this  bias just starts to exhibit a
scale dependence, both in the spatial  and velocity fields
(Eq.~\ref{eq:biases}).  In almost all cases we presented a relatively
straightforward analysis in the  main text, which was sometimes backed
up with detailed calculations in the Appendix.

We showed that, in the large scale, scale independent limit, our
expressions reduce to those of the peak background split
(Section~\ref{sec:pbs}), but in general the scale dependence  in our
model implies a much richer structure.   For example, even though the
peaks flow with the underlying field,  their velocities appear to be
biased (Section~\ref{sec:velb}).   In addition, we showed how this
$k$-dependent bias propagates into  the analysis of redshift space
distortions (Section~\ref{sec:zpk}).   We derived an exact formula for
the linear theory redshift space  correlation function
(Eq.~\ref{eq:kfisher}), and then argued  that it should be
well-approximated by a simpler expression which  has considerable
intuitive appeal (Eq.~\ref{eq:rs95correct}).   

Our formula shows that redshift space distortions of peaks can be  
modelled 
i) using the same formula (and physics) as in reference~\cite{Kaiser1987}, 
   except that various terms now become $k$-dependent, and 
ii) there is in addition a Gaussian smoothing term, which  reflects 
    the dispersion of particle velocities in linear theory.   
Thus, our formula has the same form as the phenomenological relation 
that is commonly used to model nonlinear effects, and which has been 
shown to provide increased accuracy when comparing theory with 
simulations. However, here, we demonstrated explicitly that this 
functional form is also  part and parcel of linear theory.  
Kaiser's relation \cite{Kaiser1987} assumes that the smoothing term is 
unity (the $k\ll 1$ limit) whereas Scoccimarro's formula 
\cite{Scoccimarro2004}, which is derived from the full Gaussian 
random field expression, is equivalent to expanding the Gaussian 
smoothing term and retaining only the monopole and quadrupole.  
Our result implies that linear theory can account for some of the 
effects that such a phenomenological model would otherwise ascribe 
to nonlinear evolution.

We provided a crude treatment of nonlinear effects 
(Section~\ref{sub:nonlinear}), which, though not properly accounting for 
the nonlinear evolution of the matter density and velocity fields 
\cite{CrocceScoccimarro2006b} nor for the mode-coupling contribution 
induced by  nonlinear gravitational clustering \cite{CrocceScoccimarro2006a}, 
illustrates how the new smoothing term, and the $k$-dependence of the 
(spatial and velocity) bias factors, impacts cosmological 
constraints from galaxy redshift surveys (Section~\ref{sec:cosmo}).  
Our analysis showed that allowing for a $k$-dependent velocity bias 
degrades constraints on the growth rate $f$ by at least a factor of 
two. Large cosmological simulations will be needed to ascertain 
whether dark matter halos hosting the surveyed galaxies also exhibit 
a $k$-dependent velocity bias. If they do, then improving the 
determination of $f$ will lie in our ability to model this bias.  

We also used the peaks bias model to investigate the stochasticity 
of the bias and its evolution (Section~\ref{sec:stoch}).  
We provided explicit expressions for the evolution of the scale 
dependent peaks bias and stochasticity, and argued that they helped 
to understand recent measurements of these quantities in numerical 
simulations.

As regards the evolution of peak bias, it is interesting to  consider
the peak model in light of recent work on possible  modification of
gravity.  In standard gravity, the linear theory  growth factor is
scale independent.  Therefore, a peak retains  its height when the
initial density field is linearly evolved.  However, in modified
gravity models, the linear growth factor is  $k$-dependent.  As a
result, peaks in the initial field may not  correspond to peaks in the
linearly evolved field because the shape of  the power spectra for the
two Gaussian fields is different.   This can also be seen directly by
studying the (linear theory)  motions of peaks.  In such models, the
bias of objects which  coherently flow with the matter evolves just as
it does in standard gravity \cite{HuiParfrey2008}.  Thus, whereas
objects initially placed at  maxima of the density field will still
move in accordance with  the (modified) matter flows, gravitational
motions will bring them  to positions which are no longer local maxima.

The question then arises as to whether it is the initial peaks, or
those in the evolved field, which bear a closer resemblance to the
galaxies and clusters we see today.  Presumably it is the peaks  which
have managed to survive from the initial time to the present  which
are the ones of most interest -- the ones which are transients  are
probably less interesting.  In theories with $k$-dependent  linear
growth, only the peaks with exactly the right large scale
surroundings (determined by the $k$-dependence of linear theory)  will
survive at later times; this raises the possibility that  the
correlation between galaxy clusters and their environments  can
constrain theories of large-scale modifications to gravity.   We have
not pursued this further, but note that this is consistent  with
recent analyses of dark matter halos \cite{Martinoetal2009}.

To conclude, it is worth mentioning that halo-based  approaches, which
provide a reasonably good description of the weakly nonlinear
clustering of simulated dark matter haloes and galaxies
\cite[e.g.][]{JeongKomatsu2009}, are now commonly used to extract
cosmological information  from redshift surveys (see
\cite{CooraySheth2002} for a review).  Although the dark matter halos
are the local density maxima of the evolved matter distribution, there
is no easy correspondence with the initial density maxima. This is the
reason why the peak model has somewhat fallen out of favour.  We
believe our work has shown  that many insights can be gained from a
study  of density peaks
\cite[e.g.][]{Desjacques2008b,DesjacquesSmith2008}, particularly  with
regard to a number of effects -- including scale dependence and
stochasticity of the spatial and velocity bias -- which matter in the
age of precision cosmology.

\acknowledgments

We thank the organizers of the Benasque cosmology meeting in August
2008,  where this work was initiated. We are grateful to Eiichiro
Komatsu and Roman Scoccimarro for their careful reading and comments
on the manuscript, and acknowledge useful discussions with  Adam
Amara, St\'ephane Colombi, Roman Scoccimarro, Uro\u{s} Seljak and
Robert Smith.  RKS thanks J. Bagla and the Harish Chandra Research
Institute for support during the later stages of this work, and
S. Mei, J. Bartlett and the APC, Paris 7 Diderot, where he was a
Visiting Professor when this was written up.  VD is supported by  the
Swiss National Foundation under contract No. 200021-116696/1.  RKS is
supported by NSF AST-0908241.

\appendix

\section{Checking the consistency of the peak biasing relation}
\label{sec:app1}

In this Appendix, we sketch the derivation of the cross-correlation
between peaks and the underlying density field, $\xpd(r)$, and the
averaged peak pairwise velocity, $v_{12}(r,\mu)$, which have not  been
derived previously. We compute both quantities using the peak
constraint and demonstrate that the results are consistent with  those
inferred (after a trivial calculation) from the peak biasing relation
~(\ref{eq:pkbiasing}).  See~\cite{Desjacques2008b} for complementary 
details about the calculation.

\subsection{Cross-correlation between peaks and density field}

Let $\eta_i=\partial_i\delta(\vx)/\sigma_1$ and
$\zeta_{ij}=\partial_i\partial_j\delta(\vx)/\sigma_2$ be the
normalised first and second derivatives of the density field.
Furthermore, let $\Lambda$ be the diagonal matrix of entries ${\rm
diag}(\lambda_1,\lambda_2,\lambda_3)$ where   $\lambda_1\geq
\lambda_2\geq \lambda_3$ is the non-increasing sequence of eigenvalues
of the symmetric matrix $-\zeta$.  The cross-correlation
$\xpd(r)=\la\delta\npk(\vx_1)\nu(\vx_2)\ra$ ($r=|\vx_2-\vx_1|$)
follows  from the Kac-Rice formula~\cite{KacRice},
\begin{equation}
\xpd(r) = \frac{3^{3/2}\sigma_0}{\bnpk(\nu) R_1^3}
\la|\det\zeta(\vx_1)|\delta^3[\eta(\vx_1)]\theta(\lambda_3)\nu(\vx_2)\ra\;,
\label{eq:crosspkdens}
\end{equation}
where $\bnpk$ is the differential number density of peaks of height
$\nu$ as given in Eq.~(\ref{eq:npk}), $R_1$ is the typical radius of
peaks and  $\lambda_3$ is the smallest eigenvalue of $-\zeta$ at
$\vx_1$. We have omitted the $\nu$-dependence for brevity. The
Heaviside step-function arises because we are interested in counting
maxima solely, for which  $\zeta_{ij}$ is negative definite at the
extrema position.  Unlike \cite{RegosSzalay1995} who expresses the
covariance matrix in a coordinate system where the  two density maxima
lie on the $z$-axis,  we write the expectation value in the right-hand
side of Eq.~(\ref{eq:crosspkdens}) as an integral over the angular
average,  joint probability distribution function
$P(\vy,\nu_2;r)$. Here,  $\vy^\top=(\eta_i,\nu,\zeta_A)$ is a
ten-dimensional vector whose components $\zeta_A$, $A=1,\dots,6$
symbolise the entries $ij=11,22,33,12,13,23$ of $\zeta_{ij}$.

To evaluate the 11-dimensional covariance matrix $\vcc(r)$, it is
convenient to split the degrees of freedom associated with the tensor
$\zeta$ into the scalar $u=-\tr\zeta=\sum_i\lambda_i$ (so that $u$ is
positive when $\lambda_3>0$) and the traceless matrix
$\tilde\zeta=\zeta-1/3(\tr\zeta)\vii$ , where $\vii$ is the $3\times
3$ identity matrix. Let $\tilde\zeta_A$ designate the 5 degrees of
freedom of $\tilde\zeta$. We see that $\vcc(r)=(1/4\pi)\int\!\!
d\Omega_{\rvh}\, \vcc(\vr)$ has the block diagonal decomposition
$\vcc={\rm diag}(\vcc_1,\vcc_2,\vcc_3)$. Consequently,
$P(\vy_1,\nu_2;r)$ can be expressed as a product of three joint
probability distributions
\begin{equation}
P(\vy,\nu_2;r)=P(\nu_1,u_1,\nu_2;r)\,P(\eta_1)\,P(\tilde\zeta_1)
\end{equation}
where, for shorthand convenience, subscripts denote the variables 
evaluated at different Lagrangian positions.
The 1-point probability distributions $P(\eta_1)$ and 
$P(\tilde\zeta_1)$ are 
\begin{align}
P(\eta_1) &= \left(\frac{3}{2\pi}\right)^{3/2} 
  \exp\left(-\frac{3\eta_1^2}{2}\right) \\
P(\tilde\zeta_1) &= \frac{15^3}{(2\pi)^{5/2}2\sqrt{5}}
  \exp\left[-\frac{15}{4}\tr(\tilde\zeta_1^2)\right] \nonumber \;,
\end{align}
while the joint density $P(\nu_1,u_1,\nu_2;r)$ has a covariance matrix
\begin{equation}
\vcc_1=\left(\begin{array}{ccc}
1 & \xi_0^{(0)}/\sigma_0^2 & \gamma_1\xi_0^{(1)}/\sigma_1^2 \\ 
\xi_0^{(0)}/\sigma_0^2 & 1 & \gamma_1 \\
\gamma_1\xi_0^{(1)}/\sigma_1^2 & \gamma_1 & 1 \end{array}\right)\;.
\end{equation}
Upon inversion of $\vcc_1$, the quadratic form $Q_1$ that appears in the 
probability density $P(\nu_1,u_1,\nu_2;r)$ reads as
\begin{equation}
  Q_1\left(\nu_1,u_1,\nu_2\right)=
  \frac{\nu_1^2+u_1^2-2\gamma_1\nu_1 u_1}{2\left(1-\gamma_1^2\right)}+
  \frac{\left(\nu_2-A_1\right)^2}{2\Delta_\xi}
\end{equation}
where 
\begin{align}
  \Delta_\xi &= 1-\frac{1}{\sigma_0^4}\frac{\left(\xi_0^{(0)}-
    \frac{\sigma_0}{\sigma_1}\xi_0^{(1)}\right)^2}{1-\gamma_1^2} \\
  A_1 &= \frac{1}{\sigma_0}\left[\frac{1}{\sigma_0}
    \left(\frac{\nu_1-\gamma_1 u_1}
    {1-\gamma_1^2}\right)\xi_0^{(0)} \right. \\ 
    & \qquad \left. +\frac{1}{\sigma_2}\left(\frac{u_1-\gamma_1\nu_1}
    {1-\gamma_1^2}\right)\xi_0^{(1)}\right] \nonumber\;.
\end{align}
The calculation now proceeds along lines similar to \cite{Desjacques2008b}.

We choose a coordinate system whose axes are aligned with the principal
frame of $\zeta_1$ and introduce the asymmetry parameters 
\begin{align}
v &= \left(\lambda_1-\lambda_3\right)/2 \nonumber \\  
w &= \left(\lambda_1-2 \lambda_2+\lambda_3\right)/2 \;.
\label{eq:newset}
\end{align}
Our choice of ordering impose the constraints $v\geq 0$ and  $-z\leq
w\leq v$, while the peak constraint enforces $(u+w)\geq 3v$.  Upon
integration over the angular variables that define the orientation of
the orthonormal triad of $\zeta_1$ and the variables $v$ and $w$, the
cross-correlation $\xpd(r)$ of peaks of height $\nu$ is given by
\begin{align}
  \xpd(r) &= \sigma_0G_0\left(\gamma_1,\gamma_1\nu_1\right)^{-1}
  \int_{-\infty}^{+\infty}\!\!d\nu_2\,\nu_2 
  \frac{e^{-(\nu_2-A_1)^2/2\Delta_\xi}}
       {\sqrt{2\pi \Delta_\xi}} \nonumber \\ 
       & \quad \times \int_0^\infty\!\!du_1 f(u_1)
       \frac{e^{-(u_1-\gamma_1\nu_1)^2/
	   2(1-\gamma_1^2)}}{\sqrt{2\pi\left(1-\gamma_1^2\right)}}\;.
\label{eq:xpd1}
\end{align}
where the auxiliary function $f(u)$ is defined as in Eq.~(A15) of
\cite{Bardeenetal1986}, and
\begin{equation}
G_n\!\left(\gamma,\omega\right)=\int_0^\infty\!\!\dd x\,x^n f(x)
\frac{e^{-(x-\omega)^2/2(1-\gamma^2)}}{\sqrt{2\pi\left(1-\gamma^2\right)}}\;.
\label{eq:gk}
\end{equation}
are moments of the peak curvature. In particular,
$\bar{u}(\nu)=G_1(\gamma_1,\gamma_1\nu)/G_0(\gamma_1,\gamma_1\nu)$ is
the average curvature of peaks of height $\nu$. Finally, the integral
over $\nu_2$ is performed and we arrive at the desired result:  
\begin{equation}
\xpd(r)=\frac{1}{\sigma_0}\frac{(\nu-\gamma_1\bar{u})}{(1-\gamma_1^2)}
\xi_0^{(0)}(r)+\frac{1}{\sigma_2}\frac{(\bar{u}-\gamma_1\nu)}{(1-\gamma_1^2)}
\xi_0^{(1)}(r)\;.
\label{eq:xpd2}
\end{equation}
This agrees with Eq.~(\ref{eq:xpd}), which was obtained with much less 
effort from the peak biasing relation~(\ref{eq:pkbiasing}). It is worth
noticing  that, while the derivation based on the peak biasing
relation is formally exact at  the first order only, this appendix
shows that Eq.~(\ref{eq:xpd2}) is exact to all orders.

The cross-correlation $\xi_{{\rm pk},\delta}(r)$ has a straightforward
interpretation: it is the average density profile around density
maxima, i.e. $\la \delta(\vx_2)|{\rm peak~at~}\vx_1\ra$. As shown by
\cite{Bardeenetal1986}, this constrained density profile can be
calculated easily and, after some algebra, one indeed finds $\xi_{{\rm
pk},\delta}(r)=\la \delta(\vx_2)|{\rm peak~at~}\vx_1\ra$.  Note that
$\psi(r)$ in Eq.~(7.10) of reference \cite{Bardeenetal1986}
corresponds to our $\xi_0^{(0)}(r)$.  Therefore, their 
Eq.~(7.10) appears to have an additional factor of 1/3 which 
multiplies the factors of $\nabla^2\xi_0^{(0)}=-\xi_0^{(1)}$ 
in our expression -- but this is only because they measure
$r$ in units of $R_1$ -- there is, in fact, no difference.  

\subsection{Mean streaming of peak pairs}

The calculation of the mean streaming is more involved since we have
three more degrees of freedom and an extra angular dependence. Our 
derivation is based on reference~\cite{Desjacques2008b}, who calculated 
the pairwise velocity dispersion along the line of sight.

We introduce the normalised velocity field
$\varpi_i=\mbox{v}_i/(aHf\sigma_{-1})=v_i/\sigma_{-1}$. Also, we
assume that  the line of sight axis coincides with the third-axis,
such that  $\Delta \varpi_z$ denotes  the difference
$\varpi_3(\vx_2)-\varpi_3(\vx_1)$.
The line of sight pairwise velocity weighted over all peak pairs with 
comoving separation $r$ can be expressed as
\begin{align}
\label{eq:pairwise}
\lefteqn{\left[1+\xpk(r)\right]v_{12}(r,\mu)=\bnpk^{-2}\,\sigma_{-1}} 
\\ & \quad \times \frac{1}{2\pi}\int_0^{2\pi}\!\!d\phi\,d\vy_1 d\vy_2
\Delta\varpi_z \npk(\vx_1)\npk(\vx_2) P(\vy_1,\vy_2;\vr) \nonumber \;,
\end{align}
where $\mu$ is the cosine of the angle between $\rvh=\vr/r$ and the
third axis, and $\phi$ is the azimuthal angle in the plane
perpendicular to the line of sight. The local peak density $\npk(\vx)$ 
is given by $3^{3/2}R_1^{-3}|\det\zeta(\vx)|\delta^3[\eta(\vx)]$, 
supplemented by the appropriate  conditions to select those maxima 
with a certain threshold height. 

In the above expression, the joint probability density
$P(\vy_1,\vy_2;\vr)$ is now a function of
$\vy^\top=(\varpi_i,\eta_i,\nu,\zeta_A)$, where $\eta_i\equiv 0$ owing
to the peak constraint.  The corresponding covariance matrix $\vcc$
can be decomposed into four $13\times 13$ block matrices, the
zero-point contribution $\vmm$ in the top left and bottom right
corners, and the cross-correlation matrix  $\vbb(\vr)$ and its
transpose in the bottom left and top right corners, respectively. In
the large distance limit $r\gg 1$ where $|\vbb|\ll\vmm$, an expansion
in the small perturbation $\vmm$ yields at first order
\begin{align}
\label{eq:firstorder}
P(\vy_1,\vy_2,\vr) &\approx \frac{1}{(2\pi)^{13}|\det\vcc|^{1/2}}
\left(1+\vy_1^\top\vmm^{-1}\vbb\,\vmm^{-1}\vy_2\right) \nonumber \\
& \quad \times e^{-\bar{Q}(\vy_1,\vy_2)}\;,
\end{align}
where the quadratic form $\bar{Q}(\vy_1,\vy_2)$ reads
\begin{align}
2\bar{Q}&=\frac{3\vw_1^2}{1-\gamma_0^2}+\nu_1^2
+\frac{\left(\gamma_1\nu_1+\tr\zeta_1\right)^2}{1-\gamma_1^2} \nonumber \\
& \quad +\frac{5}{2}\left[3\tr(\zeta_1^2)-\left(\tr\zeta_1\right)^2\right] 
+ 1\leftrightarrow 2\;,
\label{eq:qform}
\end{align} 
$\vw_1$ being the velocity vector at comoving position $\vx_1$.
The inverse $\vmm^{-1}$ and $\vbb(\vr)$ can be further decomposed into 
the block matrices
\begin{equation}
\vmm^{-1}=\left(\begin{array}{cc}\vpp & \vrr^\top \\  \vrr &
\vqq\end{array}\right),~~~ \vbb=\left(\begin{array}{cc} \vbb_1 &
\vbb_4^\top \\ \vbb_3 & \vbb_2 \end{array}\right) \;.
\end{equation} 
where
\begin{widetext}
\begin{equation}
\vpp=\left(\begin{array}{ccc} 
\frac{3}{(1-\gamma_0^2)}\vii & \frac{-3\gamma_0}{1-\gamma_0^2}\vii 
& 0_{3\times 1} \\
\frac{-3\gamma_0}{1-\gamma_0^2}\vii & \frac{3}{1-\gamma_0^2}\vii 
& 0_{3\times 1} \\
0_{1\times 3} & 0_{1\times 3} & (1-\gamma_1^2)^{-1} \end{array}\right), 
\quad
\vqq=\left(\begin{array}{cccccc}
\frac{6-5\gamma_1^2}{1-\gamma_1^2} & -\frac{(3-5\gamma_1^2)}{2(1-\gamma_1^2)}
& -\frac{(3-5\gamma_1^2)}{2(1-\gamma_1^2)} & 0 & 0 & 0 \\
-\frac{(3-5\gamma_1^2)}{2(1-\gamma_1^2)} & \frac{6-5\gamma_1^2}{1-\gamma_1^2}
& -\frac{(3-5\gamma_1^2)}{2(1-\gamma_1^2)} & 0 & 0 & 0 \\
-\frac{(3-5\gamma_1^2)}{2(1-\gamma_1^2)} & -\frac{(3-5\gamma_1^2)}
{2(1-\gamma_1^2)} & \frac{6-5\gamma_1^2}{1-\gamma_1^2} & 0 & 0 & 0 \\
0 & 0 & 0 & 15 & 0 & 0 \\
0 & 0 & 0 & 0 & 15 & 0 \\
0 & 0 & 0 & 0 & 0 & 15 \end{array}\right) \;,
\end{equation}
and
\begin{equation}
\vrr = \left(\begin{array}{ccc}
0_{3\times 3} & 0_{3\times 3} & \frac{\gamma_1}{1-\gamma_1^2} 1_{3\times 1} \\
0_{3\times 3} & 0_{3\times 3} & 0_{3\times 1} \end{array}\right)\;.
\end{equation}
The explicit expressions for $\vbb_i$ are too long to be given here as
they depend upon the correlation functions of $\varpi_i$, $\eta_i$,
$\nu$ and $\zeta_A$ in a rather complicated way. Fortunately,  the
mean streaming involves only the azimuthal average
$\tilde{\vbb}(r,\mu)=1/(2\pi)\int\!\!d\phi\,\vbb(\vr)$, which can 
generally be expanded as
\begin{equation}
\tilde{\vbb}(r,\mu)=\sum_{\ell=0}^4\tilde{\vbb}^\ell(r)L_\ell(\mu), \quad
\tilde{\vbb}^\ell(r)=\left(\begin{array}{cc}
\tilde{\vbb}_1^\ell & \tilde{\vbb}_4^{\ell\top} \\
\tilde{\vbb}_3^\ell & \tilde{\vbb}_2^\ell \end{array}\right)\;.
\end{equation}
Note that the multipole matrices $\tilde{\vbb}_3^\ell$ and
$\tilde{\vbb}_4^{\ell\top}$ are not independent of each other since we
have $\tilde{\vbb}_3^\ell=(-1)^\ell\tilde{\vbb}_4^\ell$.  As we will
see shortly, all the contributions but that from the $\ell=1$
multipole (unsurprisingly) cancel out. We will thus detail the results
for the dipole contribution solely. After some algebra, we find
\begin{equation}
\tilde{\vbb}_1^1(r)=\left(\begin{array}{ccccccc}
0 & 0 & 0 & 0 & 0 & 0 & 0 \\ 
0 & 0 & 0 & 0 & 0 & 0 & 0 \\ 
0 & 0 & 0 & 0 & 0 & 0 & \frac{\xi_1^{(-1/2)}}{\sigma_0\sigma_{-1}} \\ 
0 & 0 & 0 & 0 & 0 & 0 & 0 \\ 
0 & 0 & 0 & 0 & 0 & 0 & 0 \\ 
0 & 0 & 0 & 0 & 0 & 0 & \frac{\xi_1^{(1/2)}}{\sigma_0\sigma_1} \\ 
0 & 0 & -\frac{\xi_1^{(-1/2)}}{\sigma_0\sigma_{-1}} & 0 & 0 
& -\frac{\xi_1^{(1/2)}}{\sigma_0\sigma_1} & 0 \end{array}\right),\quad
\tilde{\vbb}_4^1(r)=\left(\begin{array}{ccccccc}
0 & 0 & \frac{\xi_1^{(1/2)}}{5\sigma_{-1}\sigma_2} & 0 & 0 
& \frac{\xi_1^{(3/2)}}{5\sigma_1\sigma_2} & 0 \\
0 & 0 & \frac{\xi_1^{(1/2)}}{5\sigma_{-1}\sigma_2} & 0 & 0 
& \frac{\xi_1^{(3/2)}}{5\sigma_1\sigma_2} & 0 \\
0 & 0 & \frac{3\xi_1^{(1/2)}}{5\sigma_{-1}\sigma_2} & 0 & 0 
& \frac{3\xi_1^{(3/2)}}{5\sigma_1\sigma_2} & 0 \\
0 & 0 & 0 & 0 & 0 & 0 & 0 \\ 
\frac{\xi_1^{(1/2)}}{5\sigma_{-1}\sigma_2} & 0 & 0 
& \frac{\xi_1^{(3/2)}}{5\sigma_1\sigma_2} & 0 & 0 & 0 \\
0 & \frac{\xi_1^{(1/2)}}{5\sigma_{-1}\sigma_2} & 0 & 0 
& \frac{\xi_1^{(3/2)}}{5\sigma_1\sigma_2} & 0 & 0 \end{array}\right)\;,
\end{equation}
whereas the matrix $\tilde{\vbb}_2^1$ is identically zero. Right and left 
multiplication by $\vmm^{-1}$ then gives
\begin{equation}
\vmm^{-1}\tilde{\vbb}_1^1\vmm^{-1}=\left(\begin{array}{ccccccc}
0 & 0 & 0 & 0 & 0 & 0 & 0 \\ 
0 & 0 & 0 & 0 & 0 & 0 & 0 \\ 
0 & 0 & 0 & 0 & 0 & 0 & -3\alpha_1 \\
0 & 0 & 0 & 0 & 0 & 0 & 0 \\ 
0 & 0 & 0 & 0 & 0 & 0 & 0 \\ 
0 & 0 & 0 & 0 & 0 & 0 & -3\alpha_2 \\
0 & 0 & 3\alpha_1 & 0 & 0 & 3\alpha_2 & 0 \end{array}\right),\quad
\vmm^{-1}\tilde{\vbb}_4^1\vmm^{-1}=\left(\begin{array}{ccccccc}
0 & 0 & -3\gamma_1\alpha_1 & 0 & 0 & -3\gamma_1\alpha_2 & 0 \\ 
0 & 0 & -3\gamma_1\alpha_1 & 0 & 0 & -3\gamma_1\alpha_2 & 0 \\ 
0 & 0 & 9\alpha_3-3\gamma_1\alpha_1 & 0 & 0 
& 9\alpha_4-3\gamma_1\alpha_2 & 0 \\
0 & 0 & 0 & 0 & 0 & 0 & 0 \\ 
9\alpha_3 & 0 & 0 & 9\alpha_4 & 0 & 0 & 0 \\ 
0 & 9\alpha_3 & 0 & 0 & 9\alpha_4 & 0 & 0\end{array}\right)
\end{equation}
where the functions $\alpha_i(r)$ are identical to those defined in 
Eq.~(A12) of \cite{Desjacques2008b}. Namely,
\begin{gather}
\alpha_1(r) = \frac{ \frac{\sigma_0^2}{\sigma_2^2}\xi_1^{(3/2)}
  +\frac{\sigma_0^2}{\sigma_1^2}\xi_1^{(1/2)}
  -\frac{\sigma_1^2}{\sigma_2^2}\xi_1^{(1/2)} -\xi_1^{(-1/2)}}
      {\sigma_{-1}\sigma_0\left(1-\gamma_0^2\right)
	\left(1-\gamma_1^2\right)}, \quad
      \alpha_3(r) = \frac{\xi_1^{(1/2)}-\frac{\sigma_0^2}{\sigma_1^2}
	\xi_1^{(3/2)}}{\sigma_{-1}\sigma_2\left(1-\gamma_0^2\right)} 
      \nonumber \\
\alpha_2(r) = \frac{-\frac{\sigma_1^2}{\sigma_2^2}\xi_1^{(3/2)}
  +\frac{\sigma_0^2\sigma_1^2}{\sigma_{-1}^2\sigma_2^2}\xi_1^{(1/2)}
  -\xi_1^{(1/2)} +\frac{\sigma_0^2}{\sigma_{-1}^2}\xi_1^{(-1/2)}}
      {\sigma_1\sigma_0\left(1-\gamma_0^2\right)
	\left(1-\gamma_1^2\right)}, \quad
      \alpha_4(r) = \frac{\xi_1^{(3/2)}-\frac{\sigma_0^2}
	{\sigma_{-1}^2}\xi_1^{(1/2)}}
	    {\sigma_1\sigma_2\left(1-\gamma_0^2\right)}\;.
\end{gather}
\end{widetext}
The rest of the calculation is easily accomplished owing to the
separability of the one-point probability distribution $P(\vy)$ into
the product $P_\varpi(\varpi_i) P_{\nu\zeta}(\nu,\zeta_A)$, where
$P_{\nu\zeta}$ is the one-point distribution of the density and its
second derivatives (The first derivatives merely contributes a
normalisation factor), and
\begin{equation}
P_\varpi(\varpi_i)=\frac{3^{3/2}}{(2\pi)^{3/2}\left(1-\gamma_0^2\right)^{3/2}}
\exp\left[-\frac{3\vw^2}{2\left(1-\gamma_0^2\right)}\right]
\end{equation}
is the velocity distribution of density peaks. In particular, the
first moment vanishes while the second moment $\la\varpi_i^2\ra$ is
the one-dimensional velocity dispersion of density maxima,
\begin{equation}
\la\varpi_i^2\ra=\frac{1}{3}\left(1-\gamma_0^2\right)\;.
\end{equation}
The scalar $\vy_1^\top\vmm^{-1}\tilde{\vbb}\vmm^{-1}\vy_2$ contains
terms  linear and quadratic in $\varpi_i$ as well as terms independent
of the  velocity. Upon multiplication by $\Delta\varpi_z$ and
integration over  the velocities, only quadratic terms survive. We 
eventually find
\begin{align}
\lefteqn{\int\!\!d^3\vw_1 d^3\vw_2\Delta\varpi_z\left(\vy_1^\top\vmm^{-1}
\tilde{\vbb}\vmm^{-1}\vy_2\right) P_\varpi(\vw_1) P_\varpi(\vw_2)} 
\nonumber \\
&=\left[\alpha_1\left(\nu_1+\nu_2\right)+\gamma_1\alpha_1\left(\tr\zeta_1
+\tr\zeta_2\right)-3\alpha_3\left(\zeta_{1,3}+\zeta_{2,3}\right)\right]
\nonumber \\ 
& \qquad\times \left(1-\gamma_0^2\right) L_1(\mu) \nonumber \\
& \quad -\frac{3}{2}\left[\zeta_{1,1}+\zeta_{2,1}+\zeta_{1,2}+\zeta_{2,2}
-2\left(\zeta_{1,3}+\zeta_{2,3}\right)\right] \nonumber \\
& \qquad\times \frac{\xi_3^{(1/2)}-\frac{\sigma_0^2}{\sigma_1^2}\xi_3^{(3/2)}}
{\sigma_{-1}\sigma_2} L_3(\mu)\;.
\end{align}
Here, $\zeta_{1,A}$ and $\zeta_{2,A}$ designate the component
$\zeta_A$ of  the hessian $\zeta$ at location $\vx_1$ and $\vx_2$,
respectively.  As we can see, although the even multipoles cancel out,
a term proportional to $L_3(\mu)$ remains. Also, the dipole receives a
contribution from  $-3\alpha_3(\zeta_{1,3}+\zeta_{2,3})$ which is not
invariant under rotations. However, we have to remember that the
principal axes of the tensors $\zeta_1=\zeta(\vx_1)$ and
$\zeta_2=\zeta(\vx_2)$ are not necessarily aligned with those of the
coordinate frame. Let us first consider $\zeta_1$.  Without loss of
generality, we can write $\zeta_1=-\voo\Lambda\voo^\top$, where $\voo$
is an orthogonal matrix and $\Lambda$ is the diagonal matrix
consisting of the three ordered eigenvalues $\lambda_i$ of $-\zeta_1$.
The properties of the trace implies that $\tr\zeta_1=-\tr\Lambda$, 
while $\zeta_{1,j}=-\sum_i\lambda_i\voo_{ji}^2$. Since the one-point
probability density $P(\vy)$ does not depend on $\voo$, the integral
over the SO(3) manifold that describes the orientation of the 
orthonormal triad of $\zeta_1$ is immediate,
\begin{equation}
\int\!\! d\voo\,\zeta_{1,j}=-\sum_{i=1}^3\lambda_i\int\!\! d\voo\,
\voo_{ji}^2=-\frac{1}{3}\sum_{i=1}^3\lambda_i=\frac{1}{3}\tr\zeta_1\;.
\end{equation}
Similarly, averaging over the orientation of the eigenvectors of 
$\zeta_2$ yields $\int\!\! d\voo\,\zeta_{2,j}=(1/3)\tr\zeta_2$. 
Consequently, the $\ell=3$ term vanishes and we only need to integrate
\begin{equation}
\Bigl[\alpha_1\left(\nu_1+\nu_2\right)+\left(\gamma\alpha_1-\alpha_3
\right)\left(\tr\zeta_1+\tr\zeta_2\right)\Bigr]
\left(1-\gamma_0^2\right) L_1(\mu)
\end{equation}
over the eigenvalues of $\zeta_1$ and $\zeta_2$ subjects to the peak
constraint. Substituting the expressions ~(\ref{eq:biases}) of the
bias parameters $b_\nu$ and $b_\zeta$, the result can be recast into
the form of Eq.~(\ref{eq:v12pk}) when $\nu_1=\nu_2=\nu$.

\section{The cross-correlation coefficient for powerlaw spectra}
\label{sec:app2}

In this Appendix, we examine the large-scale behaviour of the 
cross-correlation coefficient for density peaks assuming a power law
spectrum of density fluctuations.

The cross-correlation coefficient $r_\xi$ in configuration space can
be written as
\begin{equation}
\label{eq:rxlargescale}
r_\xi^2(r)=\frac{1}{1+{\cal R}(r)},~~~ 
{\cal R}(r)=\frac{\xi_0^{(2)}/\xi_0^{(0)}-\left[\xi_0^{(1)}/\xi_0^{(0)}
\right]^2}{\left[b_\nu/b_\zeta+\xi_0^{(1)}/\xi_0^{(0)}\right]^2}\;. 
\end{equation}
It is larger than unity when  ${\cal R}<0$, i.e. when
$\xi_0^{(2)}/\xi_0^{(0)}<[\xi_0^{(1)}/\xi_0^{(0)}]^2$. In Cold Dark
Matter cosmologies, the correlation functions $\xi_0^{(n)}$ must be
calculated numerically  because the spectral index is a smooth
function of wavenumber. For a no-wiggle powerlaw power spectrum
$P_\delta(k)\equiv A_s k^{n_s}$  however, they take the exact form
\begin{equation}
\label{eq:xiplaw}
\xi_0^{(n)}(r)=\frac{A_s}{4\pi^2}\,R_S^{-2\alpha}\,
\Gamma\left(\alpha\right)
\,{}_1 F_1\!\left(\alpha,\gamma;-z\right)\;,
\end{equation}
where $\Gamma(\alpha)$ and ${}_1 F_1 (\alpha,\gamma;-z)$ are the Gamma
and confluent hypergeometric function in the arguments
$\alpha=n+3/2+n_s/2$, $\gamma=3/2$, and $z=r^2/(4 R_S^2)$; and $R_S$
is the characteristic radius of the window function assumed
Gaussian. Since Eq.  (\ref{eq:rxlargescale}) only holds at large
separation $r\gg 1$, we consider the limit $|z|\to\infty$ to the above
expression, in which ${}_1 F_1 (\alpha,\gamma;-z)$ has the following
asymptotic expansion (in a suitable domain of the complex plane
\cite{GradRyz}),
\begin{widetext}
\begin{align}
\label{eq:1f1asym}
{}_1 F_1 (\alpha,\gamma;-z) &= \frac{\Gamma(\gamma)}{\Gamma(\alpha)}
e^{-z}(-z)^{\alpha-\gamma}\left(1+\sum_{k=1}^\infty (-1)^k
\frac{\Gamma(k+\gamma-\alpha)\Gamma(k+1-\alpha)}
{k!\Gamma(\gamma-\alpha)\Gamma(1-\alpha)}z^{-k}\right) \\ & \quad
+\frac{\Gamma(\gamma)}{\Gamma(\gamma-\alpha)}z^{-\alpha}
\left(1+\sum_{k=1}^\infty\frac{\Gamma(k+\alpha)\Gamma(k+\alpha-\gamma+1)}
{k!\Gamma(\alpha)\Gamma(\alpha-\gamma+1)}z^{-k}\right) \nonumber\;.
\end{align}
\end{widetext}
In the right half-plane of the variable $z$ (i.e. for Re$(z)>0$), the
first term in the right-hand side of Eq.(\ref{eq:1f1asym}) is the
subordinate part of the asymptotics, whereas the second is the
dominant part. We will now calculate ${\cal R}(r)$ for a few integer
values of the spectral index covering the range $[-3,2]$ shown in
Fig. \ref{fig:rcplaw}. As we will see shortly, the cross-correlation
coefficient  can exceed unity even if the underlying power spectrum
$P(k)$ is a featureless power law. The exact amount of stochasticity,
however, critically depends upon the shape of the underlying power
spectrum.

In the particular case of a white noise spectrum, $n_s=0$, the
dominant part cancels out owing to the fact that
$\Gamma(\gamma-\alpha)$ has simple poles at
$\gamma-\alpha=n=0,1,2,\cdots$, i.e. $\Gamma(-n)^{-1}=0$. Moreover,
$\Gamma(k+\gamma-\alpha)/\Gamma(\gamma-\alpha)=
(k-1+\gamma-\alpha)(k-2+\gamma-\alpha)\times\cdots\times(\gamma-\alpha)$
vanishes when $k\geq n+1$, so the summation in the subordinate part
involves a few terms solely.  In fact, the asymptotic expansion gives
the exact result,
\begin{gather}
\xi_0^{(0)}\!(z)=\frac{A_s}{8\pi^{3/2}R_S^3}\,e^{-z}\nonumber \\
\xi_0^{(1)}\!(z)=\frac{1}{R_S^2}\,\xi_0^{(0)}\!(z)
\left(\frac{3}{2}-z\right) \\
\xi_0^{(2)}\!(z)=\frac{1}{R_S^4}\,\xi_0^{(0)}\!(z)
\left(\frac{15}{4}-5z+z^2\right)\nonumber\;,
\end{gather}
which yields
\begin{equation}
{\cal R}(r)=\frac{\left(3-\frac{r^2}{R_S^2}\right)}
{2\left[\frac{b_\nu R_S^2}{b_\zeta}+\frac{3}{2}-\frac{r^2}{4R_S^2}\right]^2}\;.
\end{equation}
As can be seen, ${\cal R}(r)$ becomes negative at separation
$r>\sqrt{3}R_S$,  so the cross-correlation coefficient is greater than
unity at large scales.  Note, however, that the dominant part is
nonzero for any small $n_s$ different from zero. More precisely, upon
writing $n_s=\epsilon$ where $0<|\epsilon|\ll 1$ and momentarily
ignoring a factor of $A_s/(4\pi^2)R_S^{-2\alpha}$, we have
\begin{equation}
\xi_0^{(n)}\!(z)\approx \frac{\Gamma\left(n+\frac{3}{2}\right)\Gamma(3/2)}
{\Gamma\left(-n-\frac{\epsilon}{2}\right)}\,z^{-n-\frac{3}{2}}\;.
\end{equation}
The sign of ${\cal R}(r)$ is equal to that of
\begin{align}
\lefteqn{\xi_0^{(2)}/\xi_0^{(0)}-\left[\xi_0^{(1)}/\xi_0^{(0)}\right]^2}  \\
& \qquad \approx \frac{15}{4z^2}\frac{\Gamma\left(-2-\frac{\epsilon}{2}\right)}
{\Gamma\left(-\frac{\epsilon}{2}\right)}-\frac{3}{2z^2}
\left[\frac{\Gamma\left(-1-\frac{\epsilon}{2}\right)}
{\Gamma\left(-\frac{\epsilon}{2}\right)}\right]^2 \nonumber \\
& \qquad \geq 0 \nonumber\;,
\end{align}
which is positive for any small nonzero $\epsilon$. Therefore, the
cross-correlation coefficient is discontinuous at $n_s=0$. The same
analysis also shows there is a similar discontinuity at $n_s=2$.
These discontinuity points are marked as empty symbols  in
Fig.~\ref{fig:rcplaw}.

For $n_s=-2$, the dominant part is non-vanishing only when
$n=0$. Furthermore,  for  $n=1$ and 2, the subordinate part only sums
a finite number of terms.  Explicitly,
\begin{gather}
\xi_0^{(0)}\!(z)\approx\frac{A_s}{8\pi R_S}\,z^{-1/2},\quad
\xi_0^{(1)}\!(z)=\frac{A_s}{8\pi^{3/2}R_S^3}\,e^{-z} \\ 
\xi_0^{(2)}\!(z)=\frac{3A_s}{16\pi^{3/2}R_S^5}\,e^{-z}
\left(1-\frac{2z}{3}\right)\nonumber\;.
\end{gather}
On inserting these expressions into Eq.(\ref{eq:rxlargescale}), we find
\begin{equation}
{\cal R}(r)\approx \frac{3\sqrt{\pi}}{4}
\frac{\left(\frac{r}{R_S}-\frac{r^3}{6R_S^3}\right)e^{-r^2/4R_S^2}}
{\left[\sqrt{\pi}\frac{b_\nu R_S^2}{b_\zeta}+\frac{r}{2R_S}\,
e^{-r^2/4R_S^2}\right]^2}\;.
\end{equation}
Again, $r_\xi>1$ at sufficiently large separation $r\gg 1$. Note,
however, that ${\cal R}$ decays much more rapidly to zero when
$n_s=-2$.  Furthermore, one can show that $r_\xi<1$ for
$n_s=-2+\epsilon$, and $r_\xi>1$ for $n_s=-2-\epsilon$, where
$0<\epsilon\ll 1$. In other words, there is a jump discontinuity at
$n_s=-2$.

When the spectral index is an odd integer, e.g. $n_s=-3,\pm 1$, the
subordinate, complex-valued part is exponentially suppressed  relative
to the dominant, real-valued part. For $n_s=-1$, we find
\begin{gather}
\xi_0^{(0)}\!(z)\approx \frac{A_s}{8\pi^2R_S^2}\,
z^{-1}\left(1+\frac{1}{2z}\right) \\
\xi_0^{(1)}\!(z)\approx -\frac{A_s}{16\pi^2R_S^4}\,
z^{-2}\left(1+\frac{3}{z}\right) \\
\xi_0^{(2)}\!(z)\approx \frac{3 A_s}{16\pi^2R_S^6}\,
z^{-3}\left(1+\frac{15}{2z}\right)
\end{gather}
upon including the first two terms of the dominant part. After some
simplification, we arrive at
\begin{equation}
{\cal R}(r)\approx \frac{20}{r^4}
\left[\frac{b_\nu}{b_\zeta}-\frac{2}{r^2}\right]^{-2}\;.
\end{equation}
Similarly, we obtain
\begin{equation}
{\cal R}(r)\approx \frac{216}{r^4}
\left[\frac{b_\nu}{b_\zeta}-\frac{12}{r^2}\right]^{-2}
\end{equation}
for $n_s=-1$.  In both cases, ${\cal R}>0$ so the cross-correlation
coefficient is less than unity at large scales. Finally, for $n_s=-3$,
the density correlation $\xi_0^{(0)}(r)$ diverges owing to the presence
of $\Gamma(\alpha)=\Gamma(n)$. Consequently, the cross-correlation
coefficient is unity at all scales.


\begin{thebibliography}{100}

\bibitem[\protect\citeauthoryear{}{}]{distortions}{
M.~Davis, P.J.E.~Peebles, \ApJ {\bf 267}, 465 (1983);
P.B.~Lilje, G.~Efstathiou, \MNRAS {\bf 236}, 851 (1989);
J.A.~Peacock, S.J.~Dodds, \MNRAS {\bf 267}, 1020 (1994);
A.N.~Taylor, A.J.S.~Hamilton, \MNRAS {\bf 282}, 767 (1996);
W.E.~Ballinger, J.A.~Peacock, A.F.~Heavens, \MNRAS {\bf 282}, 877 (1996);
J.~Loveday, G.~Efstathiou, S.J.~Maddox, B.A.~Peterson, \ApJ {\bf 468}, 1
(1996);
A.F.~Heavens, S.~Matarrese, L.~Verde, \MNRAS {\bf 301}, 797 (1998);
H.~Magira, Y.P.~Jing, Y.~Suto, \ApJ {\bf 528}, 30 (2000);
X.~Kang, Y.P.~Jing, H.J.~Mo, G.~B\"orner, \MNRAS {\bf 336}, 892 (2002);
V.~Desjacques, A.~Nusser, \MNRAS {\bf 351}, 1395 (2004);
X.~Wang, W.~Hu, \ApJ {\bf 643}, 585 (2006);
R.E.~Smith, R.K.~Sheth, R.~Scoccimarro, \PRD {\bf 78}, 023523 (2008);
J.R.~Shaw, A.~Lewis, \PRD {\bf 78}, 103512 (2008).
\label{distortions}}

\bibitem[\protect\citeauthoryear{Kaiser}{1987}]{Kaiser1987} N.~Kaiser,
\MNRAS {\bf 227}, 1 (1987).

\bibitem[\protect\citeauthoryear{Fisher}{1995}]{Fisher1995}
K.B.~Fisher, \ApJ {\bf 448}, 494 (1995).

\bibitem[\protect\citeauthoryear{Ohta, Kayo \&
Taruya}{2004}]{Ohtaetal2004} Y. Ohta, I. Kayo, A. Taruya, \ApJ {\bf
608}, 647 (2004).

\bibitem[\protect\citeauthoryear{Peebles}{1980}]{Peebles1980}
P.J.E~Peebles, The Large-Scale Structure of the Universe (Princeton
University Press, 1980).

\bibitem[\protect\citeauthoryear{}{}]{grdeviations}{
A.~Lue, R.~Scoccimarro, G.~Starkman, \PRD, {\bf 69}, 124015 (2004);
E.V.~Linder, \PRD, {\bf 70}, 023511 (2004);
L.~Knox, Y.-S. Song, J.A.~Tyson, \PRD, {\bf 74}, 023512 (2006);
M.~Ishak, A.~Upadhye, D.~Spergel, \PRD, {\bf 74}, 043513 
(2006).\label{grdeviations}}

\bibitem[\protect\citeauthoryear{D\"urrer \& Maartens}{2008}]
{DurrerMaartens2008} R.~D\"urrer, R.~Maartens, arXiv:00811.4132 (2008).

\bibitem[\protect\citeauthoryear{Hamilton}{1992}]{Hamilton1992}
A.J.S. Hamilton, \ApJL {\bf 385}, L5 (1992).

\bibitem[\protect\citeauthoryear{Cole, Fisher \&
Weinberg}{1995}]{Coleetal1995} S. Cole, K.B. Fisher, D. Weinberg,
1995, \MNRAS {\bf 275}, 515 (1995).

\bibitem[\protect\citeauthoryear{Scoccimarro}{2004}]{Scoccimarro2004}
R.~Scoccimarro, \PRD {\bf 70}, 083007 (2004).

\bibitem[\protect\citeauthoryear{Matsubara}{1999}]{Matsubara1999}
T.~Matsubara, \ApJ {\bf 525}, 543 (1999).

\bibitem[\protect\citeauthoryear{Doroshkevich}{1970}]{Doroshkevich1970}
A.G.~Doroshkevich, Astrofizika {\bf 3}, 175 (1970).

\bibitem[\protect\citeauthoryear{Gorski}{1988}]{Gorski1988} K. Gorski,
\ApJL {\bf 332}, L7 (1988).

\bibitem[\protect\citeauthoryear{Kaiser}{1984}]{Kaiser1984} N.~Kaiser,
\ApJ {\bf 284}, L9 (1984).

\bibitem[\protect\citeauthoryear{Peacock \&
Heavens}{1985}]{PeacockHeavens1985} J.A.~Peacock, A.F.~Heavens,
\MNRAS {\bf 217}, 805 (1985).

\bibitem[\protect\citeauthoryear{Hoffman \& Shaham}{1985}]
{HoffmanShaham1985} Y.~Hoffman, J.~Shaham, \ApJ {\bf 297}, 16 (1985).

\bibitem[\protect\citeauthoryear{Bardeen et
al.}{1986}]{Bardeenetal1986}  J.M.~Bardeen, J.R.~Bond, N.~Kaiser,
A.S.~Szalay, \ApJ {\bf 304}, 15  (1986).

\bibitem[\protect\citeauthoryear{Coles}{1989}]{Coles1989} P.~Coles,
\MNRAS {\bf 238}, 319 (1989).

\bibitem[\protect\citeauthoryear{Lumsden, Heavens \& Peacock}{1989}]
{Lumsdenetal1989} S.L.~Lumsden, A.F.~Heavens, J.A.~Peacock, \MNRAS
{\bf 238}, 293 (1989).

\bibitem[\protect\citeauthoryear{Reg\"{o}s \&
Szalay}{1995}]{RegosSzalay1995} E.~Reg\"{o}s. A.S.~Szalay, \MNRAS
{\bf 272}, 447 (1995).

\bibitem[\protect\citeauthoryear{Bond \& Myers}{1996}]{BondMyers1996}
J.R.~Bond, S.T.~Myers, \ApJS {\bf 103}, 1 (1996).

\bibitem[\protect\citeauthoryear{Sheth, Mo \&
Tormen}{2001}]{ShethMoTormen2001} R.K. Sheth, H.J. Mo, G. Tormen,
\MNRAS {\bf 323}, 1 (2001).

\bibitem[\protect\citeauthoryear{Desjacques}{2008}]{Desjacques2008a}
V.~Desjacques, \MNRAS {\bf 388}, 638 (2008).

\bibitem[\protect\citeauthoryear{Desjacques \& Smith}{2008}]
{DesjacquesSmith2008} V.~Desjacques, R.E.~Smith, \PRD {\bf 78},
023527 (2008).

\bibitem[\protect\citeauthoryear{Kaiser \&
Davis}{1985}]{KaiserDavis1985} N.~Kaiser, M.~Davis, \ApJ {\bf 297},
365 (1985).

\bibitem[\protect\citeauthoryear{Mo, Jing \&
White}{1997}]{MoJingWhite1997} H.J. Mo, Y.P. Jing, S.D.M. White,
\MNRAS {\bf 284}, 189 (1997).

\bibitem[\protect\citeauthoryear{Cen}{1998}]{Cen1998} R.~Cen, \ApJ
{\bf 509}, 494 (1998).

\bibitem[\protect\citeauthoryear{Sheth}{2001}]{Sheth2001} R.K. Sheth, 
Annals of the New York Academy of Sciences {\bf 927}, 1 (2001).

\bibitem[\protect\citeauthoryear{Desjacques}{2008}]{Desjacques2008b}
V.~Desjacques, \PRD, {\bf 78}, 103503 (2008).

\bibitem[\protect\citeauthoryear{}{}]{biases}{In terms of the {\it
normalised} (and smoothed) variables $\nu_S=\delta_S/\sigma_0$ and
$u_S=-\nabla^2\delta_S/\sigma_2$, the peak number density is
$\delta\npk=\sigma_0 b_\nu\nu_S+\sigma_2 b_\zeta u_S$ at the first
order. This shows that the relative importance of the $b_\nu$ and
$b_\zeta$ terms is controlled by $\sigma_0 b_\nu$ and $\sigma_2
b_\zeta$.}

\bibitem[\protect\citeauthoryear{Percival \&
White}{2009}]{PercivalWhite2009} W.J.~Percival, M.~White, \MNRAS {\bf
393}, 297 (2009).

\bibitem[\protect\citeauthoryear{Gunn \& Gott}{1972}]{GunnGott1972}
J.E.~Gunn, J.R.~Gott III, \ApJ {\bf 176}, 1 (1972).

\bibitem[\protect\citeauthoryear{Press \& Schechter}{1974}]
{PressSchechter1974} W.H.~Press, P.~Schechter, \ApJ {\bf 187}, 425
(1974).

\bibitem[\protect\citeauthoryear{Komatsu et
al.}{2009}]{Komatsuetal2009} E.~Komatsu, et al., \ApJS {\bf 180}, 330
(2009).

\bibitem[\protect\citeauthoryear{Mandelbaum et
al.}{2006}]{Mandelbaumetal2006} R.~Mandelbaum, U.~Seljak, R.J.~Cool,
M.~Blanton, C.M.~Hirata, J.~Brinkmann,  \MNRAS {\bf 372}, 758 (2006).

\bibitem[\protect\citeauthoryear{Kulkarni et
al.}{2007}]{Kulkarnietal2007} G.~Kulkarni et al., \MNRAS {\bf 378},
1196 (2007).

\bibitem[\protect\citeauthoryear{Wake et al.}{2008}]{Wake2dFSDSS2008}
D. Wake et al., \MNRAS {\bf 387}, 1045 (2008).

\bibitem[\protect\citeauthoryear{Szalay}{1988}]{Szalay1988}
A.S.~Szalay, \ApJ {\bf 333}, 21 (1988).

\bibitem[\protect\citeauthoryear{Fry \& Gazta\~{n}aga}{1993}]
{FryGaztanaga1993} J.N.~Fry, E.~Gazta\~{n}aga, \ApJ {\bf 413}, 447
(1993).

\bibitem[\protect\citeauthoryear{Coles}{1993}]{Coles1993}
P.~Coles, \MNRAS {\bf 262}, 1065 (1993).

\bibitem[\protect\citeauthoryear{Taruya et al.}{2001}]{Taruya2001}
A. Taruya, H. Magira, Y.P. Jing, Y. Suto, \PASJ {\bf 53}, 155 (2001).

\bibitem[\protect\citeauthoryear{Mo \& White}{1996}]{MoWhite1996}
H.J.~Mo, S.D.M.~White, \MNRAS {\bf 282}, 347 (1996).

\bibitem[\protect\citeauthoryear{Cole \&
Kaiser}{1989}]{ColeKaiser1989} S. Cole, N. Kaiser, \MNRAS {\bf 237},
1127 (1989).

\bibitem[\protect\citeauthoryear{Sheth \&
Tormen}{1999}]{ShethTormen1999} R.K. Sheth, G. Tormen, \MNRAS {\bf
308}, 119 (1999).

\bibitem[\protect\citeauthoryear{Bharadwaj}{2001}]{Bharadwaj2001}
S.~Bharadwaj, \MNRAS {\bf 327}, 577 (2001).

\bibitem[\protect\citeauthoryear{}{}]{v12error}{Our expression for 
$v_{12}$ corrects a sign error  in Eq.~(50) of \cite{Desjacques2008b},
which propagated to Fig.~8  of that paper.}

\bibitem[\protect\citeauthoryear{}{}]{mistake}{
We believe expression for the  real space peak power spectrum 
in Ref.~[21], their Eq.~(70), should read
 $\Pi(k) = P(k)\,(x+y)^2/\sigma_0^2$.  Once
corrected, this relation is equivalent to our Eq.~(\ref{eq:xpk})
provided that $\sigma_0 b_\nu=\la W|{\cal C}\ra /\sqrt{1-\gamma_1^2}$
and  $\sigma_2 b_\zeta = \la X|{\cal C}\ra -\gamma_1\la W|{\cal
C}\ra/\sqrt{1-\gamma_1^2}$ (see their paper for details about their
notation).  For the redshift space power, they have the same Gaussian
damping term  as we do, but their expression for $\ppk^{s0}$, their
Eq.~(84), does  not reduce to the square of peak density and velocity
bias terms.   In their Eq.~(84), their $w$ should be a $y$, and their
$x-y$ should  be $x+y$ (this is the same error that affected their
expression for  the real space power spectrum; it also affects their
expression for  $v_{12}$).  These errors appear to have propagated to
their Figure~6.}

\bibitem[\protect\citeauthoryear{Szalay \&
Jensen}{1987}]{SzalayJensen1987} A.S.~Szalay, L.G.~Jensen, Acta
Physica Hungarica {\bf 62}, 263 (1987).

\bibitem[\protect\citeauthoryear{Peacock et
al.}{1987}]{Peacocketal1987} J.A.~Peacock, S.L.~Lumsden, A.F.~Heavens,
\MNRAS {\bf 229}, 469 (1987).

\bibitem[\protect\citeauthoryear{Percival \& Sch\"afer}{2008}]
{PercivalSchafer2008} W.J.~Percival, B.M.~Sch\"afer, \MNRAS {\bf 385},
L78 (2008).

\bibitem[\protect\citeauthoryear{Tegmark et
al.}{2006}]{TegmarkSDSS2006} M. Tegmark et al. \PRD {\bf 74}, 123507
(2006).

\bibitem[\protect\citeauthoryear{Song \&
Percival}{2008}]{SongPercival2008} Y.-S. Song, W.J.~Percival, 
astro-ph/0807.0810 (2008).

\bibitem[\protect\citeauthoryear{Sheth}{1996}]{Sheth1996} R.K. Sheth,
\MNRAS {\bf 279}, 1310 (1996).

\bibitem[\protect\citeauthoryear{Jackson}{1972}]{Jackson1972}
J.C.~Jackson, \MNRAS {\bf 156}, 1 (1972).

\bibitem[\protect\citeauthoryear{Hamilton}{1998}]{Hamilton1998}
A.J.S.~Hamilton, ``Linear Redshift Distortions: a Review'', in ``The
evolving Universe'', ed. D.~Hamilton (Kluwer Academic Publishers,
1998).

\bibitem[\protect\citeauthoryear{Sheth \&
Diaferio}{2001}]{ShethDiaferio2001} R.K. Sheth, A. Diaferio, \MNRAS
{\bf 322}, 901 (2001).

\bibitem[\protect\citeauthoryear{Eisenstein, Seo \&
White}{2007}]{EisensteinSeoWhite07} D.J.~Eisenstein, H.-J.~Seo,
M.~White, \ApJ {\bf 664}, 660 (2007).

\bibitem[\protect\citeauthoryear{Sheth et al.}{2001}]{Shethetal2001}
R.K. Sheth, L. Hui, A. Diaferio, R. Scoccimarro,  \MNRAS {\bf 325},
1288 (2001).

\bibitem[\protect\citeauthoryear{Bharadwaj}{1996}]{Bharadwaj1996}
S.~Bharadwaj, \ApJ {\bf 460}, 28 (1996).

\bibitem[\protect\citeauthoryear{Crocce \&
Scoccimarro}{2008}]{CrocceScoccimarro08} M.~Crocce, R.~Scoccimarro,
\PRD {\bf 77}, 023533 (2008).

\bibitem[\protect\citeauthoryear{Matsubara}{2008a}]{Matsubara2008a}
T.~Matsubara, \PRD {\bf 77}, 063530 (2008a).

\bibitem[\protect\citeauthoryear{Matsubara}{2008b}]{Matsubara2008b}
T.~Matsubara, \PRD {\bf 78}, 083519 (2008b).

\bibitem[\protect\citeauthoryear{Kim et al.}{2009}]{Kimetal2009}
J.~Kim, C.~Park, R.J.~Gott, J.~Dubinski, \ApJ {\bf 701}, 1547 (2009).

\bibitem[\protect\citeauthoryear{Manera et al.}{2009}]{Maneraetal2009}
M.~Manera, R.K.~Sheth, R.~Scoccimarro, arXiv:0906.1314 (2009).

\bibitem[\protect\citeauthoryear{Peacock \&
Dodds}{1994}]{PeacockDodds1994} J.A.~Peacock, S.J.~Dodds, \MNRAS {\bf
267}, 1020 (1994).

\bibitem[\protect\citeauthoryear{White et al.}{2009}]{Whiteetal2009}
M.~White, Y.-S.~Song, W.J.~Percival, \MNRAS {\bf 397}, 1348 (2009).

\bibitem[\protect\citeauthoryear{Taylor et al.}{2001}]{Tayloretal2001}
A.N.~Taylor, W.E.~Ballinger, A.F.~Heavens, H.~Tadros, \MNRAS 
{\bf 327}, 689 (2001).

\bibitem[\protect\citeauthoryear{Nesseris \& Perivolaropoulos}{2008}]
{NesserisPerivolaropoulos2008} S.~Nesseris, L.~Perivolaropoulos, \PRD
{\bf 77}, 023504 (2008).

\bibitem[\protect\citeauthoryear{Feldman et
al.}{1994}]{Feldmanetal1994} H.A.~Feldman, N.~Kaiser, J.A.~Peacock,
\ApJ {\bf 426}, 23 (1994).

\bibitem[\protect\citeauthoryear{Vogeley \&
Szalay}{1996}]{VogeleySzalay1996} M.S.~Vogeley, A.S.~Szalay, \ApJ
{\bf 465}, 34 (1996).

\bibitem[\protect\citeauthoryear{Tegmark et
al.}{1998}]{Tegmarketal1998} M.~Tegmark, A.J.S~Hamilton, M.A.~Strauss,
M.S.~Vogeley, A.S.~Szalay, \ApJ {\bf 499}, 555 (1998).

\bibitem[\protect\citeauthoryear{Stril et al.}{2009}]{Striletal2009}
A.~Stril, R.N.~Cahn, E.V.~Linder, arXiv:0910.1833 (2009).

\bibitem[\protect\citeauthoryear{Seljak et al.}{2009}]{Seljaketal2009}
U.~Seljak, N.~Hamaus, V.~Desjacques, \PRL {\bf 103}, 091303 (2009).

\bibitem[\protect\citeauthoryear{Burkey \&
Taylor}{2004}]{BurkeyTaylor2004} D.~Burkey, A.N.~Taylor, \MNRAS {\bf
347}, 255 (2004).

\bibitem[\protect\citeauthoryear{Seljak}{2009}]{Seljak2009}
U.~Seljak, \PRL {\bf 102}, 021302 (2009).

\bibitem[\protect\citeauthoryear{McDonald \&
Seljak}{2008}]{McDonaldSeljak2008} P.~McDonald, U.~Seljak,
arXiv:0810.0323 (2008).

\bibitem[\protect\citeauthoryear{Jain \& Zhang}{2008}]{JainZhang2008}
B.~Jain, P.~Zhang, \PRD {\bf 78}, 063503 (2008).

\bibitem[\protect\citeauthoryear{Song \& Koyama}{2008}]{SongKoyama2008}
Y.-S.~Song, K.~Koyama, \JCAP {\bf 01}, 048 (2008).

\bibitem[\protect\citeauthoryear{Scherrer \& Weinberg}{1998}]
{ScherrerWeinberg1998} R.J.~Scherrer, D.H.~Weinberg, \ApJ {\bf 504}, 
607 (1998).

\bibitem[\protect\citeauthoryear{Pen}{1998}]{Pen1998}
U.-L.~Pen, \ApJ {\bf 504}, 601 (1998).

\bibitem[\protect\citeauthoryear{Dekel \& Lahav}{1999}]{DekelLahav1999}
A.~Dekel, O.~Lahav, \ApJ {\bf 520}, 24 (1999).

\bibitem[\protect\citeauthoryear{Fry}{1996}]{Fry1996} J.N.~Fry, \ApJL
{\bf 461}, L65 (1996).

\bibitem[\protect\citeauthoryear{Hui \&
Parfrey}{2008}]{HuiParfrey2008} L.~Hui, K. Parfrey, \PRD {\bf 77},
043527 (2008).

\bibitem[\protect\citeauthoryear{Tegmark \&
Peebles}{1998}]{TegmarkPeebles1998} M. Tegmark, P.J.E. Peebles, \ApJL
{\bf 500}, L79 (1998).

\bibitem[\protect\citeauthoryear{Martino et
al.}{2009}]{Martinoetal2009} M. Martino, H. F. Stabenau, R.K. Sheth,
\PRD {\bf 79}, 084013 (2009).

\bibitem[\protect\citeauthoryear{Crocce \& Scoccimarro}{2006}]{CrocceScoccimarro2006b}
M.~Crocce, R.~Scoccimarro, \PRD {\bf 73}, 063520 (2006).

\bibitem[\protect\citeauthoryear{Crocce \& Scoccimarro}{2006}]{CrocceScoccimarro2006a}
M.~Crocce, R.~Scoccimarro, \PRD {\bf 73}, 063519 (2006).

\bibitem[\protect\citeauthoryear{Cooray \&
Sheth}{2002}]{CooraySheth2002} A. Cooray, R.K. Sheth, Phys. Rep. {\bf
372}, 1 (2002).

\bibitem[\protect\citeauthoryear{Jeong \& Komatsu}{2009}]{JeongKomatsu2009}
D.~Jeong, E.~Komatsu, \ApJ {\bf 691}, 569 (2009).

\bibitem[\protect\citeauthoryear{}{}]{KacRice} {M.~Kac, \BAMS {\bf
 49}, 314 (1943); S.O.~Rice, Mathematical analysis of random noise, in
 Selected Papers on  Noise and Stochastic Processes, Dover, New York
 (1954).}

\bibitem[\protect\citeauthoryear{Gradshteyn \& Ryzhik}{1980}]{GradRyz}
I.S.~Gradshteyn, I.M.~Ryzhik, Table of Integrals, Series and Products,
6th edition (Academic Press, 2000).

\end{thebibliography}
\end{document}